\documentclass[epj]{svjour}
\usepackage{graphics}
\RequirePackage{graphicx}
\RequirePackage{multirow}
\RequirePackage{yhmath}
\RequirePackage{hyperref}
\usepackage{color}
\usepackage{cases}
\usepackage{subfigure}
\usepackage{dcolumn,booktabs,bm}
\usepackage{slashed}
\usepackage{amsfonts,amssymb,stmaryrd,latexsym,amsmath}
\usepackage{textcomp}
\usepackage{braket}
\usepackage{widetext}

\newcommand{\etal}{\textit{et al}. }

\renewcommand\tabcolsep{1.5 pt}

\def\m{\mathcal}

\begin{document}
\title{Possible Molecular States Composed of Doubly Charmed Baryons with Couple-channel Effect}
\author{Bin Yang\inst{1}\thanks{e-mail: bin{\_}yang@pku.edu.cn } \and Lu Meng\inst{1}\thanks{e-mail: lmeng@pku.edu.cn} \and Shi-Lin Zhu\inst{1,2,3}%
\thanks{e-mail: zhusl@pku.edu.cn}
}                     % Do not remove

\institute{School of Physics and State Key Laboratory of Nuclear
Physics and Technology, Peking University, Beijing 100871, China\label{addr1} \and Center of High Energy Physics, Peking University, Beijing 100871, China\label{addr2} \and  Collaborative Innovation Center of Quantum Matter, Beijing 100871, China \label{addr3}}
\date{Received: date / Revised version: date}
\abstract{
We systematically investigate the possible molecular states composed
of (1) two spin-$\frac{3}{2}$ doubly charmed baryons, and (2) a pair
of spin-$\frac{3}{2}$ and spin-$\frac{1}{2}$ doubly charmed baryons.
The one-boson-exchange (OBE) model is used to describe the potential
between two baryons. The channel mixing effect is considered for the
systems with the same quantum number $(I(J^P))$ but different total
spin ($S$) and orbital angular momenta ($L$). We also study the
channel mixing effect among the systems composed of various doubly
charmed baryons if they have the same quantum number. Many of the
systems are good candidates of molecular states.
%
%\PACS{
%      {PACS-key}{discribing text of that key}   \and
%      {PACS-key}{discribing text of that key}
%     } % end of PACS codes
} %end of abstract
\maketitle
%
%%%%%%%%%%%%%%%%%%%%
\section{Introduction}\label{sec_intro}
%%%%%%%%%%%%%%%%%%%%

The researches on exotic states attract much attention since the
first charmonium-like exotic state $X(3872)$ was reported by the
Belle Collaboration~\cite{Choi:2003ue} in 2003. After that, more and
more exotic states have been observed by many major experimental
collaborations, such as CLEO-c, BaBar, Belle, BESIII, CDF, D0, LHCb
and CMS. Those states include charmonium-like states such as
$Z_c(3900)$~\cite{Liu:2013dau,Ablikim:2013mio,Xiao:2013iha},
$Y(4260)$~\cite{Aubert:2005rm,He:2006kg},
$Y(4140)$~\cite{Aaltonen:2009tz}, and bottomonium-like states,
$Z_b(10610)$ and $Z_b(10650)$~\cite{Belle:2011aa} and so on. In
2015, the LHCb Collaboration discovered two pentaquark states
$P_c(4380)$ and $P_c(4450)$ in the $J/\Psi$ invariant mass spectrum
of the $\Lambda^0_b\rightarrow J/\Psi K^- p$
decay~\cite{Aaij:2015tga}. Recently, the LHCb Collaboration reported
that $P_c(4450)$ was resolved into two states $P_c(4440)$ and
$P_c(4457)$ and observed a lower state
$P_c(4312)$~\cite{Aaij:2019vzc}. The experimental and theoretical
progress about exotic states can be found in the recent
reviews~\cite{Chen:2016qju,Esposito:2016noz,Chen:2016spr,Lebed:2016hpi,Guo:2017jvc,Olsen:2017bmm,Yuan:2018inv,Liu:2019zoy}.

Some multiquark exotic states are near the threshold of two hadrons.
They might be molecular states.
A hadronic molecular state is a loosely bound system composed of two color-singlet hadrons.
The two hadrons are bound together by the residual force of the color interaction. One can use the One-boson-exchange (OBE) model to describe the dynamics between the two baryons in a molecular system.
The OBE model is very successful to describe the deuteron as a hadronic molecular state composed of a neutron and a proton.
The meson exchange force together with the couple-channel effect between $S$-wave and $D$-wave renders the deuteron a loosely bound state.

About forty years ago, Voloshin and Okun proposed the hadronic molecule composed of two heavy mesons ~\cite{Voloshin:1976ap}.
De Rujula \etal interpreted the $\psi(4040)$ as a $D^*\bar{D}^*$ molecule in Ref.~\cite{DeRujula:1976zlg}.
T\"ornqvist used the one-pion-exchange (OPE) potential to calculate the possible charmed meson-antimeson molecular state~\cite{Tornqvist:1993ng,Tornqvist:1993vu}.
Besides, there are also many other calculations about possible hadronic molecular states, such as the combination of two mesons~\cite{Wong:2003xk,Liu:2009ei,Close:2009ag,Ding:2009vj,Sun:2011uh,Li:2012ss,Zhao:2014gqa}, or two baryons~\cite{Lee:2011rka,Meguro:2011nr,Li:2012bt,Carames:2015sya,Meng:2017fwb,Meng:2017udf,Yang:2018amd}.
Similarly, one can also explain the hidden charm pentaquark states as a molecular state formed by one heavy meson and one heavy baryon~\cite{Wu:2010jy,Yang:2011wz,Chen:2015loa,Roca:2015dva,Yamaguchi:2017zmn,Shimizu:2017xrg,Yamaguchi:2016ote,Chen:2019asm,Meng:2019ilv}.

In 2017, the LHCb collaboration reported the doubly charmed baryon
$\Xi_{cc}$ at
$3621.40\pm0.72(\text{stat})\pm0.27(\text{syst})\pm0.14(\Lambda_c^+)\text{
	MeV}$ in the $\Lambda_c^+K^-\pi^+\pi^+$ mass
spectrum~\cite{Aaij:2017ueg}. As an important member in the baryon
family, there are many theoretical works to calculate the mass of
the doubly charmed
baryon~\cite{Gershtein:1998sx,Itoh:2000um,Zhang:2008rt,Wang:2010hs,Sun:2016wzh,Shah:2016vmd,Shah:2017liu,Weng:2018mmf}.
It is also very interesting to investigate the possible molecular
states containing doubly charmed baryons. The system composed of
$\Xi_{cc}$ and a charmed meson or baryon was calculated in
Refs.~\cite{Chen:2017jjn,Chen:2018pzd}. The possible molecular
system with $\Xi_{cc}$ and a nucleon is studied in
Refs.~\cite{Froemel:2004ea,Meng:2017udf}.

In Ref.~\cite{Meng:2017fwb}, the authors investigated the possible deuteron-like bound states composed of two spin-$\frac{1}{2}$ doubly charmed baryons in the SU(3) flavor symmetry.
In this work, we extend the same formalism to investigate the possible hadronic molecular states composed of two spin-$\frac{3}{2}$ doubly charmed baryons (include $\Xi_{cc}$ and $\Omega_{cc}$), denoted as $B^*B^*$, as well as system one spin-$\frac{3}{2}$ and one spin-$\frac{1}{2}$ baryon, denoted as $B^*B$. We use the OBE model to describe the potential between two baryons.
The couple-channel effect among $B^*B^*$, $B^*B$ and two spin-$\frac{1}{2}$ baryons, $BB$, are also included in this work.
When we calculate the pure $B^*B^*$ and $B^*B$ systems, we consider the couple-channel effect from $D$-wave and $G$-wave.

We organize this work as follows.
We give the theoretical formalism in Section~\ref{sec_form}, including the effective Lagrangian, coupling constants and the effective interaction potentials.
We present the numerical results of the $B^*B^*$ systems and the $B^*B$ systems in Section~\ref{sec_res}.
In the calculation, we also include the couple-channel effect among $BB$, $B^*B$ and $B^*B^*$.
Then we discuss our results and conclude in Section~\ref{sec_dis}.
In Appendixes~\ref{app_Four} and \ref{app_ope}, we collect some useful formulae and functions. We also calculate the systems composed of one baryon and one antibaryon, such as $B^*\bar{B}^*$ and $B^*\bar{B}$.
The results are collected in Appendix~\ref{app_BBbar}.

%%%%%%%%%%%%%%%%%%%%%%%%%%%%%%%%%%%%%%%%%%%%
\section{Formalism}\label{sec_form}
%%%%%%%%%%%%%%%%%%%%%%%%%%%%%%%%%%%%%%%%%%%%

For a doubly charmed baryon, the two charm quarks can be treated as a static color source in the heavy quark limit.
For the two charm quarks, their color wave function should be in the antisymmetric $\bar{3}_c$-representation.
The spatial wave function of the two charm quarks is symmetric for the ground state.
As a consequence, their spin wave function must be symmetric because of the Pauli principle.
Therefore, the total spin of the two charm quark is 1, and the total spin of a ground doubly charmed baryon is $1\over2$ or $3\over2$.
In the present work, we focus on the possible deuteron-like systems composed of two doubly charmed baryons with both spin-$3\over2$, or one spin-$3\over2$ baryon and one spin-$1\over2$ baryon. The systems with the same quantum number but different components may couple with each other. We include the couple-channel effect in this work.

\subsection{The Lagrangian}\label{subsec_lag}
For convenience, we use column matrices to describe the doubly charmed baryons as follows,
\begin{equation}
B=\left[ \begin{array}{ccc}
\Xi_{cc}^{+} & \Xi_{cc}^{++} & \Omega_{cc}^{+}\end{array}\right]^{T},~~
B^{*\mu}=\left[ \begin{array}{ccc}
\Xi_{cc}^{*+} & \Xi_{cc}^{*++} & \Omega_{cc}^{*+}\end{array}\right]^{\mu T},
\end{equation}
where $B$ and $B^{*\mu}$ denote the spin-$1\over2$ and spin-$3\over2$ baryons, respectively. The exchanged mesons between two baryons are denoted by two matrices as follows,
\begin{eqnarray}
\mathcal{M} &=& \left[
\begin{array}{ccc}
\frac{\pi^0}{\sqrt{2}}+\frac
{\eta}{\sqrt{6}}&     \pi^+               &  K^+  \\
\pi^-                   &-\frac{\pi^0}{\sqrt{2}}+\frac{\eta}{\sqrt{6}}&  K^0   \\
K^-                  &\bar{K}^0                    &-\frac{2}{\sqrt{6}}\eta  \\
\end{array}
\right], \nonumber
\\
\mathcal{V}^{\mu} &=& \left[
\begin{array}{ccc}
\frac{\rho^0}{\sqrt{2}}+\frac{\omega}{\sqrt{2}}&     \rho^{+}               &  K^{*+}  \\
\rho^-                    &-\frac{\rho^0}{\sqrt{2}}+\frac{\omega}{\sqrt{2}} &  K^{*0}   \\
K^{*-}                  &\bar{K}^{*0}                    & \phi  \\
\end{array}
\right]^{\mu},
\end{eqnarray}
where $\mathcal{M}$ and $\mathcal{V}^{\mu}$ denote the octet pseudoscalar mesons and the nonet vector mesons, respectively.

We construct the effective Lagrangian as follows,
\begin{equation}
\m{L}=\mathcal{L}_{\sigma hh}+\mathcal{L}_{phh}+\mathcal{L}_{vhh},
\end{equation}
for the scalar meson exchange
\begin{equation}\label{L_s}
\m{L}_{\sigma hh}=g_{\sigma BB}\bar{B}\sigma B-g_{\sigma B^{*}B^{*}}\bar{B}^{*\mu}\sigma B_{\mu}^{*},
\end{equation}
for the pseudoscalar meson exchange
\begin{eqnarray}\label{L_ps}
\begin{aligned}
\m{L}_{phh}=&-\frac{g_{pBB}}{2m_{B}}\bar{B}\gamma_{\mu}\gamma_{5}\partial^{\mu}\mathcal{M}B+\frac{g_{pB^{*}B^{*}}}{2m_{B^{*}}}\bar{B}^{*\mu}\gamma_{\nu}\gamma_{5}\partial^{\nu}\mathcal{M}B_{\mu}^{*}
\\
&+\frac{g_{pB^{*}B}}{m_{B}+m_{B^{*}}}\bar{B}_{\mu}^{*}\partial^{\mu}\mathcal{M}B+h.c.,
\end{aligned}
\end{eqnarray}
and for the vector meson exchange
\begin{eqnarray}\label{L_v}
\begin{aligned}
\m{L}_{vhh}=&g_{vBB}\bar{B}\gamma_{\mu}\mathcal{V}^{\mu}B+\frac{f_{vBB}}{2m_{B}}\bar{B}\sigma_{\mu\nu}\partial^{\mu}\mathcal{V}^{\nu}B
\\
&-g_{vB^{*}B^{*}}\bar{B}^{*\mu}\gamma_{\nu}\mathcal{V}^{\nu}B_{\mu}^{*}
\\
&-i\frac{f_{vB^{*}B^{*}}}{2m_{B^{*}}}\bar{B}_{\mu}^{*}(\partial^{\mu}\mathcal{V}^{\nu}-\partial^{\nu}\mathcal{V}^{\mu})B_{\nu}^{*}
\\
&+i\frac{f_{vB^{*}B}}{2\sqrt{m_{B^{*}}m_{B}}}\bar{B}_{\mu}^{*}(\partial^{\mu}\mathcal{V}^{\nu}-\partial^{\nu}\mathcal{V}^{\mu})\gamma_{\nu}\gamma_{5}B+h.c..
\end{aligned}
\end{eqnarray}
The notations $g_{\sigma B^{(*)}B^{(*)}}$, $g_{pB^{(*)}B^{(*)}}$,
$g_{vB^{(*)}B^{(*)}}$ and $f_{vB^{(*)}B^{(*)}}$, represent the
coupling constants.
$m_B$ and $m_{B^*}$ are the masses of the spin-$\frac{1}{2}$ and spin-$\frac{3}{2}$ heavy baryons, respectively.

\begin{table*}[htp]
	\centering
	\caption{The relevant hadron masses~\cite{Aaij:2017ueg,Weng:2018mmf,Ebert:2002ig} and coupling constants for the nucleon~\cite{Machleidt:1987hj,Riska:2000gd,Machleidt:2000ge,Cao:2010km}. For the multiple hadrons, their averaged masses are used.}\label{Table_mass}
	\begin{tabular}{cc|cc|cc|cc}
		\hline
		Baryons & Mass( MeV) & Mesons & Mass( MeV) & Mesons & Mass( MeV) & Couplings & Value\tabularnewline
		\hline
		$\Xi_{cc}$ & 3621.4 & $\sigma$ & 600 & $\rho$ & 775.49 & $g_{\sigma NN}^{2}/4\pi$ & 5.69\tabularnewline
		$\Omega_{cc}$ & 3778 & $\pi$ & 137.25 & $\omega$ & 782.65 & $g_{\pi NN}^{2}/4\pi$ & 13.6\tabularnewline
		$\Xi_{cc}^{*}$ & 3686 & $\eta$ & 547.85 & $\phi$ & 1019.46 & $g_{\rho NN}^{2}/4\pi$ & 0.84\tabularnewline
		$\Omega_{cc}^{*}$ & 3872 & $K$ & 495.65 & $K^{*}$ & 893.77 & $f_{\rho NN}/g_{\rho NN}$ & 6.1\tabularnewline
		\hline
	\end{tabular}
\end{table*}

\subsection{Coupling constants}\label{subsec_coupling}

The coupling constants in the effective
Lagrangian~(\ref{L_s}-\ref{L_v}) should be extracted from the
experiment data. However there are no experiment data for the doubly
charmed baryon scattering with light mesons. Thus, we compare the
coupling constants of doubly charmed baryons with the ones of the
nucleons, which are known. With the help of the quark model, we get
the relations between the coupling constants of doubly charmed
baryons and nucleons.
The details of this method can be found in Ref.~\cite{Riska:2000gd}.
Here we directly show the relations between the two sets of coupling constants,
\begin{itemize}
	\item scalar meson exchange
	\begin{equation}\label{cc_s}
	g_{\sigma BB}=g_{\sigma B^{*}B^{*}}=\frac{1}{3}g_{\sigma NN},
	\end{equation}
	\item pseudoscalar meson exchange
	\begin{equation}
	g_{pBB}=-\frac{\sqrt{2}}{5}\frac{m_{B}}{m_{N}}g_{\pi NN},
	\end{equation}
	\begin{equation}
	g_{pB^{*}B^{*}}=\frac{3\sqrt{2}}{5}\frac{m_{\Xi_{cc}^{*}}}{m_{N}}g_{\pi NN},
	\end{equation}
	\begin{equation}
	g_{pB^{*}B}=\frac{2\sqrt{6}}{5}\frac{m_{B}+m_{B^{*}}}{2m_{N}}g_{\pi NN},
	\end{equation}
	\item vector meson exchange
	
	\begin{equation}\label{cc_v1}
	\begin{split}
	g_{vBB}&=\sqrt{2}g_{\rho NN}
	\\
	g_{vBB}+f_{vBB}&=-\frac{\sqrt{2}}{5}(g_{\rho NN}+f_{\rho NN})\frac{m_{B}}{m_{N}},
	\end{split}
	\end{equation}
	
	\begin{equation}\label{cc_v2}
	\begin{split}
	g_{vB^{*}B^{*}}&=\sqrt{2}g_{\rho NN}
	\\
	g_{vB^{*}B^{*}}+f_{vB^{*}B^{*}}&=\frac{3\sqrt{2}}{5}(g_{\rho NN}+f_{\rho NN})\frac{m_{B^{*}}}{m_{N}},
	\end{split}
	\end{equation}
	
	\begin{equation}\label{cc_v3}
	\begin{split}
	g_{vB^{*}B}&=0
	\\
	f_{vB^{*}B}&=-\frac{2\sqrt{6}}{5}\frac{\sqrt{m_{B^{*}}m_{B}}}{m_{N}}(g_{\rho NN}+f_{\rho NN}).
	\end{split}
	\end{equation}
\end{itemize}
In the above formula, $g_{\sigma NN}$, $g_{\pi NN}$ $g_{\rho NN}$ and $f_{\rho NN}$ are the corresponding coupling constants of nucleons.
The values of them are taken from Refs.~\cite{Machleidt:1987hj,Riska:2000gd,Machleidt:2000ge,Cao:2010km}.
In this work, we choose three nuclear coupling constants as input.
The coupling constant $g_{\pi NN}$ is stable in various models compared with $g_{\omega NN}$ and $g_{\eta NN}$. Therefore, we use the unified parameter of the nucleon-nucleon-$\pi$ vertex for the pseudoscalar meson exchange.
For the vector meson exchange, we choose the parameter of the $\rho$ meson exchange vertex for the same reason.
Their values are given in Table~\ref{Table_mass}. We give the values of the coupling constants for doubly charmed baryons in Table~\ref{Table_coupling}.

\begin{table*}[htp]
	\centering
	\caption{The coupling constants for the doubly charmed baryons.}\label{Table_coupling}
	\begin{tabular}{ccccc|ccccc|ccc}
		\hline
		& $g_{\sigma BB}$ & $g_{pBB}$ & $g_{vBB}$ & $f_{vBB}$ &  & $g_{\sigma B^{*}B^{*}}$ & $g_{pB^{*}B^{*}}$ & $g_{vB^{*}B^{*}}$ & $f_{vB^{*}B^{*}}$ &  & $g_{pB^{*}B}$ & $f_{vB^{*}B}$\tabularnewline
		\hline
		$\Xi_{cc}\Xi_{cc}$ & 2.82 & -14.26 & 4.60 & -29.76 & $\Xi_{cc}^{*}\Xi_{cc}^{*}$ & 2.82 & 43.55 & 4.60 & 72.25 & $\Xi_{cc}^{*}\Xi_{cc}$ & 49.84 & -87.95\tabularnewline
		$\Xi_{cc}\Omega_{cc}$ & 2.82 & -14.57 & 4.60 & -30.30 & $\Xi_{cc}^{*}\Omega_{cc}^{*}$ & 2.82 & 44.65 & 4.60 & 74.16 & $\Xi_{cc}^{*}\Omega_{cc}$ & 50.91 & -89.83\tabularnewline
		$\Omega_{cc}\Omega_{cc}$ & 2.82 & -14.88 & 4.60 & -30.85 & $\Omega_{cc}^{*}\Omega_{cc}^{*}$ & 2.82 & 45.75 & 4.60 & 76.12 & $\Xi_{cc}\Omega_{cc}^{*}$ & 51.11 & -90.14\tabularnewline
		&  &  &  &  &  &  &  &  &  & $\Omega_{cc}^{*}\Omega_{cc}$ & 52.18 & -92.07\tabularnewline
		\hline
	\end{tabular}
\end{table*}

\subsection{The effective potentials}\label{subsec_potential}

We give the effective Lagrangian in Section~\ref{subsec_lag}.
One can write the effective potentials in the momentum space for two doubly charmed baryons with the Lagrangians~(\ref{L_s}-\ref{L_v}).
The effective potentials $V(Q)$ is a function of the momentum of the exchanged mesons.
In the non-relativistic limit, we reserve the scattering amplitude up to $\m{O}(\frac{1}{m_Q^2})$, where $m_Q$ is the mass of the doubly charmed baryon.
Given that the baryons are not fundamental point particles, we introduce a form factor $\m F(\bm{Q})$ at each baryon-baryon-meson vertex,
\begin{equation}
\m{F}(\textbf{Q})=\frac{\Lambda^{2}-m_{ex}^{2}}{\Lambda^{2}-Q^{2}}=\frac{\Lambda^{2}-m_{ex}^{2}}
{\lambda^{2}+\textbf{Q}^{2}},
\end{equation}
where $\lambda^2=\Lambda^{2}-Q_0^{2}$.
$m_{ex}$ is the mass of the exchanged mesons.
The parameter $\Lambda$ is an adjustable momentum cutoff, which can suppress the contribution of the high momentum transferred.
We use the factor to roughly describe the effect of the baryon structure.
As a low energy effective model, the OBE potential should not contain very short-range interactions.
Therefore, adopting the form factor keeps the model self-consistent.
The parameter $\Lambda$ can not be strictly determined without any experiment data.
The value of $\Lambda$ is around an empirical scale, 1 GeV.
The possible binding energy of a two baryon system depends on the parameter, as we shall show in Section~\ref{sec_res}.
Then we transform the effective potential together with the form factor to coordinate space,
\begin{equation}
V(r)=\frac{1}{(2\pi)^3}\int d\vec{Q}e^{i\vec{Q}\cdot\vec{r}}V(\vec{Q})\m{F}^2(\vec{Q}).
\end{equation}

The total effective potential is as follows,
\begin{equation}\label{eff_pot}
V(r)=V^{s}(r)+V^{p}(r)+V^{v}(r).
\end{equation}
The superscripts $s$, $p$ and $v$ are used to mark the scalar, pseudoscalar and vector mesons exchange potentials.
The potentials can be divided into four terms.
They are the central potential term, the spin-spin interaction term, the spin-orbit interaction term and the tensor term.
The spin-spin, spin-orbit and tensor terms contain the spin-spin operator $\Delta_{SS}$, spin-orbit operator $\Delta_{LS}$ and tensor operator $\Delta_{T}$, respectively.
The specific definition of those angular momentum dependent operators are different for the systems composed of baryons with different spins.
We will discuss the operators in detail after we introduce the couple-channel effect.
Here we use subscripts $C$, $SS$, $LS$, $T$ to mark them respectively.
We derive the general potentials between two doubly charmed baryons in the specific formulae for different mesons exchanged as follows.
\begin{itemize}
	\item For the scalar meson exchange,
	\begin{equation}\label{V_S}
	\begin{split}
	V^{s}(r)=&V_{C}^{s}(r,\sigma)+V_{LS}^{s}(r,\sigma),\\
	V_{C}^{s}(r)=&-C_{\sigma}^{s}\frac{g_{\sigma1}g_{\sigma2}}{4\pi}u_{\sigma}\left[H_{0}-\frac{u_{\sigma}^{2}}{8m_{a}m_{b}}H_{1}\right],
	\\
	V_{LS}^{s}(r)=&-C_{\sigma}^{s}\frac{g_{\sigma1}g_{\sigma2}}{4\pi}\frac{u_{\sigma}^{3}}{2m_{a}m_{b}}H_{2}\Delta_{LS}.
	\end{split}
	\end{equation}
	\item For the pseudoscalar meson exchange,
	\begin{equation}\label{V_P1}
	\begin{split}
	V^{p}(r)&=\underset{\alpha=\pi, \eta, K}{\sum}\left[V_{SS}^{p}(r,\alpha)+V_{T}^{p}(r,\alpha)\right],\\
	V_{SS}^{p}(r,\alpha)&=C_{\alpha}^{p}\frac{g_{p1}g_{p2}}{4\pi}\frac{u_{\alpha}^{3}}{12m_{a}m_{b}}H_{1}\Delta_{SS},
	\\
	V_{T}^{p}(r,\alpha)&=C_{\alpha}^{p}\frac{g_{p1}g_{p2}}{4\pi}\frac{u_{\alpha}^{3}}{12m_{a}m_{b}}H_{3}\Delta_{T}.
	\end{split}
	\end{equation}
	if $u_{\alpha}^2<0$, the potentials change into
	\begin{equation}\label{V_P2}
	\begin{split}
	V_{SS}^{p}(r,\alpha)&=C_{\alpha}^{p}\frac{g_{p1}g_{p2}}{4\pi}\frac{\theta_{\alpha}^{3}}{12m_{a}m_{b}}M_{1}\Delta_{SS},
	\\
	V_{T}^{p}(r,\alpha)&=C_{\alpha}^{p}\frac{g_{p1}g_{p2}}{4\pi}\frac{\theta_{\alpha}^{3}}{12m_{a}m_{b}}M_{3}\Delta_{T},
	\end{split}
	\end{equation}
	where $\theta_{\alpha}^2=-u_{\alpha}^2$.
	\item For the vector meson exchange,
	\begin{widetext}
		\begin{equation}\label{V_V}
		\begin{split}
		V^{v}(r)=&\underset{\beta=\omega, \rho, \phi, K^*}{\sum}\left[V_{C}^{v}(r,\beta)+V_{SS}^{v}(r,\beta)+V_{T}^{v}(r,\beta)+V_{LS}^{v}(r,\beta)\right],
		\\
		V_{C}^{v}(r,\beta)=&C_{\beta}^{v}\frac{u_{\beta}}{4\pi}\bigg[g_{v1}g_{v2}H_{0}+\frac{u_{\beta}^{2}}{8m_{a}m_{b}}
		(g_{v1}g_{v2}+2g_{v1}f_{v2}+2g_{v2}f_{v1})H_{1}\bigg],
		\\
		V_{SS}^{v}(r,\beta)=&C_{\beta}^{v}\frac{1}{4\pi}\bigg[g_{v1}g_{v2}+g_{v1}f_{v2}+g_{V2}f_{v1}+f_{v1}f_{v2}\bigg]
		\frac{u_{\beta}^{3}}{6m_{a}m_{b}}H_{1}\Delta_{SS},
		\\
		V_{T}^{v}(r,\beta)=&-C_{\beta}^{v}\frac{1}{4\pi}\bigg[g_{v1}g_{v2}+g_{v1}f_{v2}+g_{v2}f_{v1}+f_{v1}f_{v2}\bigg]
		\frac{u_{\beta}^{3}}{12m_{a}m_{b}}H_{3}\Delta_{T},
		\\
		V_{LS}^{v}(r,\beta)=&-C_{\beta}^{v}\frac{1}{4\pi}\bigg[3g_{v1}g_{v2}\Delta_{LS}+4g_{v1}f_{v2}\Delta_{LS_{a}}
		+4g_{v2}f_{v1}\Delta_{LS_{b}}\bigg]\frac{u_{\beta}^{3}}{2m_{a}m_{b}}H_{2}.
		\end{split}
		\end{equation}
	\end{widetext}
\end{itemize}

In the above expressions, $u_{\sigma/\alpha/\beta}$ are defined as $u_{ex}^2=m_{ex}^2-Q_{0}^2$, where $m_{ex}$ is the mass of the exchanged meson.
$Q_{0}$ is the zeroth component of the transition momentum.
In the heavy baryon limit, $Q_{0}$ can be easily expressed with mass differences between the initial and final baryons.
For the system composed of two baryons with masses $m_1$ and $m_2$, the value of the $Q_{0}$ is 0 for the direct diagram. For the cross diagram, the value of the $Q_{0}$ is $|m_1-m_2|$.
$m_a$ and $m_b$ are the masses of two doubly charmed baryons.
For the initial and the final baryons of one Fermion line with the mass splitting, we approximately choose the geometric means of their masses, $m_a=\sqrt{m_{ai} m_{af}}$.
In Table~\ref{Table_mass} we give the masses of the baryons, as well as the masses of the exchanged mesons.
For the multiple hadrons, their averaged masses are used.

\begin{figure}
	\centering
	\includegraphics[width=0.618\linewidth]{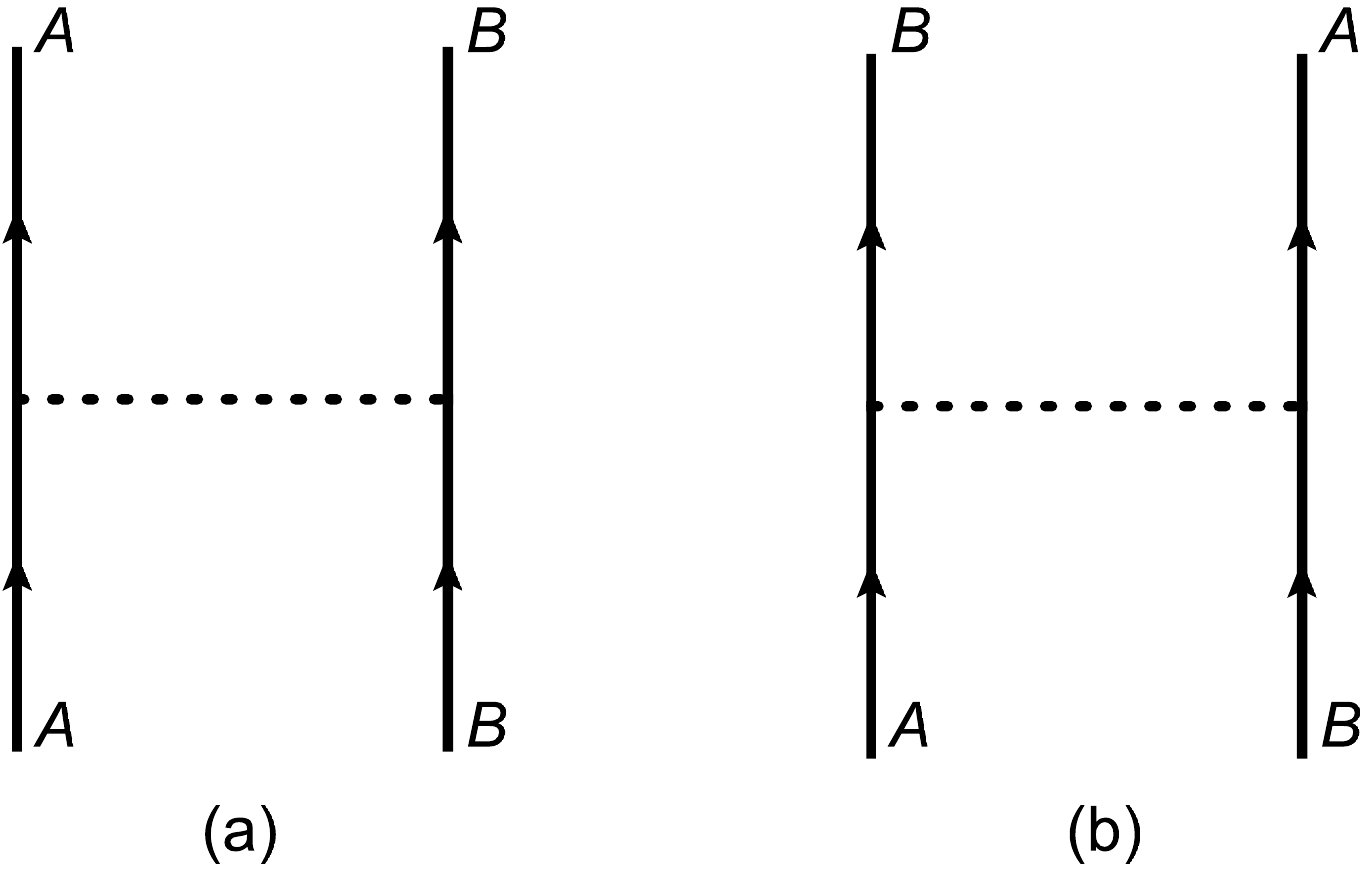}
	\caption{Feynman diagrams for the systems composed of different baryons. ``a" and ``b" are direct and cross diagrams respectively.}
	\label{fig:feynmandiagram}
\end{figure}

$g_s, g_p$ and $g_v$ are the coupling constants in Eqs.~(\ref{cc_s}-\ref{cc_v3}).
The subscript $1$ and $2$ of the coupling constants are used to mark different vertices.
$C_{\sigma}^{s}$, $C_{\alpha}^{p}$ and $C_{\beta}^{v}$ are the isospin factors.
Their values are given in Table~\ref{Table_if}.
For a system composed of different baryons, we should consider both direct and cross diagrams, as in Fig.~\ref{fig:feynmandiagram}.
In Table~\ref{Table_if} we use ``[ ]" to denote the isospin factors of the cross diagrams.
The functions $H_i=H_i(\Lambda,m_{\sigma/\alpha/\beta},r)$ and $M_i=M_i(\Lambda,m_{\alpha},r)$ come from the Fourier transformation.
We give their specific expressions in Appendix~\ref{app_Four}.

For the system composed of one baryon and one antibaryon, the Lagrangians~(\ref{L_s}-\ref{L_v}) still work.
For the process exchanging a meson with certain G-parity, $I_G$, the potentials are the Eqs.~(\ref{V_S}-\ref{V_V}) with the extra G-parity factor $I_G$.
With the help of the G-parity rule, we can directly write down the effect potentials of the baryon and antibaryon systems.
We let the isospin factors of baryon-antibaryon system absorb the extra G-parity factor.
The values are also given in Table~\ref{Table_if}.

\begin{table*}[htp]
	\centering
	\caption{The isospin factors for two baryon systems and baryon-antibaryon systems. The factors $I_G$ from G-parity rule have been absorbed by the isospin factors in the right panel.}\label{Table_if}
	\begin{tabular}{ccccccccc|ccccccccc}
		\hline
		States & $C_{\sigma}^{s}$ & $C_{\pi}^{p}$ & $C_{\eta}^{p}$ & $C_{K}^{p}$ & $C_{\rho}^{v}$ & $C_{\omega}^{v}$ & $C_{\phi}^{v}$ & $C_{K^{*}}^{v}$ & States & $C_{\sigma}^{s}$ & $C_{\pi}^{p}$ & $C_{\eta}^{p}$ & $C_{K}^{p}$ & $C_{\rho}^{v}$ & $C_{\omega}^{v}$ & $C_{\phi}^{v}$ & $C_{K^{*}}^{v}$\tabularnewline
		\hline
		$[\Xi_{cc}^{*}\Xi_{cc}^{*}]^{I=0}$ & 1 & -3/2 & 1/6 & 0 & -3/2 & 1/2 & 0 & 0 & $[\Xi_{cc}^{*}\bar{\Xi}_{cc}^{*}]^{I=0}$ & 1 & 3/2 & 1/6 & 0 & -3/2 & -1/2 & 0 & 0\tabularnewline
		$[\Xi_{cc}^{*}\Xi_{cc}^{*}]^{I=1}$ & 1 & 1/2 & 1/6 & 0 & 1/2 & 1/2 & 0 & 0 & $[\Xi_{cc}^{*}\bar{\Xi}_{cc}^{*}]^{I=1}$ & 1 & -1/2 & 1/6 & 0 & 1/2 & -1/2 & 0 & 0\tabularnewline
		$[\Xi_{cc}^{*}\Omega_{cc}^{*}]^{I=\frac{1}{2}}$ & 1 & 0 & -1/3 & 0{[}1{]} & 0 & 0 & 0 & 0{[}1{]} & $[\Xi_{cc}^{*}\bar{\Omega}_{cc}^{*}]^{I=\frac{1}{2}}$ & 1 & 0 & -1/3 & 0 & 0 & 0 & 0 & 0\tabularnewline
		$[\Omega_{cc}^{*}\Omega_{cc}^{*}]^{I=0}$ & 1 & 0 & 2/3 & 0 & 0 & 0 & 1 & 0 & $[\Omega_{cc}^{*}\bar{\Omega}_{cc}^{*}]^{I=0}$ & 1 & 0 & 2/3 & 0 & 0 & 0 & -1 & 0\tabularnewline
		\hline
	\end{tabular}
\end{table*}

For some systems, some terms in the Lagrangian are lacking.
Their potentials can be described with part of the formulae~(\ref{L_s}-\ref{L_v}).
In those cases we can also directly use the same potential but set the relevant coupling constant to zero.

We consider the couple-channel effect between states with different orbital angular momentum.
The spin of a system composed of two spin-$\frac{3}{2}$ baryons can be 0, 1, 2 and 3.
For a $J=0$ system, we consider the couple-channel effect between ${}^1S_0$ and ${}^5D_0$.
For a $J=1$ system, we consider four channels, ${}^3S_1$, ${}^3D_1$, ${}^7D_1$ and ${}^7G_1$.
For a system with total spin 2, there are also four channels, ${}^5S_2$, ${}^1D_2$, ${}^5D_2$ and ${}^5G_2$.
And there are five channels for a $J=3$ system, ${}^7S_3$, ${}^3D_3$, ${}^7D_3$, ${}^3G_3$ and ${}^7G_3$.
We list all possible channels in Table~\ref{Table_cc33}.
Their wave functions can be expressed as follows,
\begin{widetext}
	
	\begin{itemize}
		\item For a $J=0$ system,
		\begin{equation}
		\Psi(r,\theta,\phi)^{T}\chi_{ss_{z}}^{T}=
		\left[\begin{array}{c}
		T_{S}(r)\\
		0
		\end{array}\right]\ket{{}^1S_0}+
		\left[\begin{array}{c}
		0\\
		T_{D}(r)
		\end{array}\right]\ket{{}^5D_0}.
		\end{equation}
		\item For $J=1$ system,
		\begin{equation}
		\Psi(r,\theta,\phi)^{T}\chi_{ss_{z}}^{T}=
		\left[\begin{array}{c}
		T_{S}(r)\\
		0\\
		0\\
		0
		\end{array}\right]\ket{{}^3S_1}+
		\left[\begin{array}{c}
		0\\
		T_{D_1}(r)\\
		0\\
		0
		\end{array}\right]\ket{{}^3D_1}+
		\left[\begin{array}{c}
		0\\
		0\\
		T_{D_2}(r)\\
		0
		\end{array}\right]\ket{{}^7D_1}+
		\left[\begin{array}{c}
		0\\
		0\\
		0\\
		T_{G}(r)
		\end{array}\right]\ket{{}^7G_1}.
		\end{equation}
		\item For a $J=2$ system,
		\begin{equation}
		\Psi(r,\theta,\phi)^{T}\chi_{ss_{z}}^{T}=
		\left[\begin{array}{c}
		T_{S}(r)\\
		0\\
		0\\
		0
		\end{array}\right]\ket{{}^5S_2}+
		\left[\begin{array}{c}
		0\\
		T_{D_1}(r)\\
		0\\
		0
		\end{array}\right]\ket{{}^1D_2}+
		\left[\begin{array}{c}
		0\\
		0\\
		T_{D_2}(r)\\
		0
		\end{array}\right]\ket{{}^5D_2}+
		\left[\begin{array}{c}
		0\\
		0\\
		0\\
		T_{G}(r)
		\end{array}\right]\ket{{}^5G_2}.
		\end{equation}
		\item For a $J=3$ system,
		\begin{equation}
		\Psi(r,\theta,\phi)^{T}\chi_{ss_{z}}^{T}=
		\left[\begin{array}{c}
		T_{S}(r)\\
		0\\
		0\\
		0\\
		0
		\end{array}\right]\ket{{}^7S_3}+
		\left[\begin{array}{c}
		0\\
		T_{D_1}(r)\\
		0\\
		0\\
		0
		\end{array}\right]\ket{{}^3D_3}+
		\left[\begin{array}{c}
		0\\
		0\\
		T_{D_2}(r)\\
		0\\
		0
		\end{array}\right]\ket{{}^7D_3}+
		\left[\begin{array}{c}
		0\\
		0\\
		0\\
		T_{G_1}(r)\\
		0
		\end{array}\right]\ket{{}^3G_3}+
		\left[\begin{array}{c}
		0\\
		0\\
		0\\
		0\\
		T_{G_2}(r)
		\end{array}\right]\ket{{}^7G_3}.
		\end{equation}
	\end{itemize}
\end{widetext}
In the expression, $\Psi(r,\theta,\phi)$ and $\chi_{ss_{z}}$ are the spatial and spin wave functions, respectively.

The spin of a system composed of one spin-$\frac{3}{2}$ baryon and one spin-$\frac{1}{2}$ baryon can be 1 and 2.
For a $J=1$ system, we consider the channels mixing effect between ${}^3S_1$, ${}^3D_1$ and ${}^5D_1$.
For a system with total spin 2, there are four channels should be considered, ${}^5S_2$, ${}^3D_2$, ${}^5D_2$ and ${}^5G_2$.
We also list them in Table~\ref{Table_cc33}.
Their wave functions are the same.

\begin{table}[h]
	\centering
	\caption{The channels considered in this work for two baryons systems.}\label{Table_cc33}
	\begin{tabular}{c|ccccc|cccc}
		\hline
		& \multicolumn{5}{c|}{$B^{*}B^{*}$} & \multicolumn{4}{c}{$B^{*}B$}\tabularnewline
		\hline
		& $S$ & $D1$ & $D2$ & $G1$ & $G2$ & $S$ & $D1$ & $D2$ & $G$\tabularnewline
		\hline
		$J=0$ & $|^{1}S_{0}\rangle$ & $|^{5}D_{0}\rangle$ &  &  &  &  &  &  & \tabularnewline
		$J=1$ & $|^{3}S_{1}\rangle$ & $|^{3}D_{1}\rangle$ & $|^{7}D_{1}\rangle$ & $|^{7}G_{1}\rangle$ &  & $|^{3}S_{1}\rangle$ & $|^{3}D_{1}\rangle$ & $|^{5}D_{1}\rangle$ & \tabularnewline
		$J=2$ & $|^{5}S_{2}\rangle$ & $|^{1}D_{2}\rangle$ & $|^{5}D_{2}\rangle$ & $|^{5}G_{2}\rangle$ &  & $|^{5}S_{2}\rangle$ & $|^{3}D_{2}\rangle$ & $|^{5}D_{2}\rangle$ & $|^{5}G_{2}\rangle$\tabularnewline
		$J=3$ & $|^{7}S_{3}\rangle$ & $|^{3}D_{3}\rangle$ & $|^{7}D_{3}\rangle$ & $|^{3}G_{3}\rangle$ & $|^{7}G_{3}\rangle$ &  &  &  & \tabularnewline
		\hline
	\end{tabular}
\end{table}

The angular momentum dependent operators, $\Delta_{SS}$, $\Delta_{LS}$ and $\Delta_{T}$ have different forms for various combinations of the baryon spins.
We give their specific expressions in Table~\ref{Table_ope}.
We give more details about the definition of the operator matrices in Appendix~\ref{app_ope}.

\begin{table*}[htp]
	\centering
	\caption{The specific expressions of the operators $\Delta_{SS}$, $\Delta_{LS}$ and $\Delta_{T}$ for different channels.
		$\bm{\sigma}$ is the Pauli matrix.
		$\bm{S}=\frac{3}{2}\bm{\sigma}_{rs}$ is the spin operator of spin-$\frac{3}{2}$ baryons.
		$\bm{S}_t$ is the transition spin operator for the Rarita-Schwinger field.
		$\bm{L}$ is the relative orbit angular momentum operator between the two baryons.
		More details about the operator matrices are given in Appendix~\ref{app_ope}. }\label{Table_ope}
	\begin{tabular}{l|ccc}
		\hline
		Channels & $\Delta_{SS}$ & $\Delta_{LS}$ & $\Delta_{T}$\tabularnewline
		\hline
		$B^*B^*\rightarrow B^*B^*$ & $\bm{\sigma}_{rsA}\cdot\bm{\sigma}_{rsB}$ & $\frac{1}{2}\bm{L}\cdot\bm{\sigma}_{rs}$ & $3\bm{\sigma}_{rsA}\cdot\hat{\bm{r}}\bm{\sigma}_{rsB}\cdot\hat{\bm{r}}-\bm{\sigma}_{rsA}\cdot\bm{\sigma}_{rsB}$\tabularnewline
		
		$B^*B\rightarrow B^*B$ & $\bm{\sigma}_{rsA}\cdot\bm{\sigma}_{B}$ &
		$\frac{1}{2}\bm{L}\cdot\left(\bm{\sigma}_{rsA}+\bm{\sigma}_{B}\right)$ & $3\bm{\sigma}_{rsA}\cdot\hat{\bm{r}}\bm{\sigma}_{B}\cdot\hat{\bm{r}}-\bm{\sigma}_{rsA}\cdot\bm{\sigma}_{B}$\tabularnewline
		
		$BB^*\rightarrow B^*B$ & $\bm{S}_{tA}^{\dagger}\cdot\bm{S}_{tB}$ & $\frac{1}{2}\bm{L}\cdot\left(\bm{S}_{tA}^{\dagger}+\bm{S}_{tB}\right)$ & $3\bm{S}_{tA}^{\dagger}\cdot\hat{\bm{r}}\bm{S}_{tB}\cdot\hat{\bm{r}}-\bm{S}_{tA}^{\dagger}\cdot\bm{S}_{tB}$\tabularnewline

		$B^*B\rightarrow B^*B^*$ & $\bm{\sigma}_{rsA}\cdot\bm{S}_{tB}^{\dagger}$ & $\frac{1}{2}\bm{L}\cdot\left(\bm{\sigma}_{rsA}+\bm{S}_{tB}^{\dagger}\right)$ & $3\bm{\sigma}_{rsA}\cdot\hat{\bm{r}}\bm{S}_{tB}^{\dagger}\cdot\hat{\bm{r}}-\bm{\sigma}_{rsA}\cdot\bm{S}_{tB}^{\dagger}$\tabularnewline

		$B^*B\rightarrow BB$ & $\bm{S}_{tA}\cdot\bm{\sigma}_{B}$ & $\frac{1}{2}\bm{L}\cdot\left(\bm{S}_{tA}+\bm{\sigma}_{B}\right)$ & $3\bm{S}_{tA}\cdot\hat{\bm{r}}\bm{\sigma}_{B}\cdot\hat{\bm{r}}-\bm{S}_{tA}\cdot\bm{\sigma}_{B}$\tabularnewline

		$BB\rightarrow BB$ & $\bm{\sigma}_{A}\cdot\bm{\sigma}_{B}$ & $\frac{1}{2}\bm{L}\cdot\bm{\sigma}$ & $3\bm{\sigma}_{A}\cdot\hat{\bm{r}}\bm{\sigma}_{B}\cdot\hat{\bm{r}}-\bm{\sigma}_{A}\cdot\bm{\sigma}_{B}$\tabularnewline
		
		\hline
	\end{tabular}
\end{table*}

%%%%%%%%%%%%%%%%%%%%%%%%%%%%%%%%%%%%%%%%%%%%
\section{Numerical results}\label{sec_res}
%%%%%%%%%%%%%%%%%%%%%%%%%%%%%%%%%%%%%%%%%%%%

With the potential in Eq.~\eqref{eff_pot}, we solve the Schr\"odinger equation numerically.
For the system with a binding solution, we give the binding energy ($B.E.$).
The wave function of a bound state solution can also be used to check whether it is rational to treat the system as a molecular state.
Using the wave function, we give the root-mean-square radius ($R_{rms}$),
\begin{equation}
R_{rms}^2=\int \sum_i T^*_i(r)T_i(r) r^4dr,
\end{equation}
where $T_i$ is the radial wave function of the $i$th channel.
Meanwhile, we can get the individual probability for each channel,
\begin{equation}
P_{T_i}=\int T_i^*(r)T_i(r)r^2dr.
\end{equation}

In our calculation, the cutoff parameter $\Lambda$ cannot be determined exactly without relevant experiment data.
We treat it as a free parameter and choose different values in a reasonable range for different systems.
The experience on the study of the deuteron with one-boson-exchange-potential gives us some advice.
It is very successful to describe the deuteron with the cutoff between 0.8 GeV and 1.5 GeV.
Since the two doubly charmed baryon system is much heavier than the
deuteron, we also consider the cutoff up to 2.5 GeV.

%%%%%%%%%%%%%%%%%%%%%%%%%%%%%%%%%%%%%%%%%%%%
\subsection{Two spin-$\frac{3}{2}$ baryons system}\label{subsec_res1}
%%%%%%%%%%%%%%%%%%%%%%%%%%%%%%%%%%%%%%%%%%%%

For the $\Xi_{cc}^*\Xi_{cc}^*$ and $\Omega_{cc}^*\Omega_{cc}^*$
systems, we should consider the Pauli Principle. For example, the
total isospin for the $\Omega_{cc}^*\Omega_{cc}^*$ system is
symmetric, the total spin for the system can only be 0 and 2. For
the $\Xi_{cc}^*\Omega_{cc}^*$ systems, all combinations of spin and
isospin are possible. We show the binding energies and the
root-mean-square radii of possible molecular states in
Table~\ref{Table_B3B3}. Because of the tensor operator, the channels
with the same total angular momentum but different spin and orbital
angular momenta mix each other. We also show the percentage of the
different channels in the Table.

\begin{table*}[htb]
	\centering
	\caption{The numerical results for the $B^*B^*$ systems. $\Lambda $ is the cutoff parameter. ``$B.E.$'' is the binding energy. $R_{rms}$ is the root-mean-square radius. $P_{i}$ is the percentage of the different channels.}\label{Table_B3B3}
	\begin{tabular}{ccccccccc}
		\hline
		States & $\Lambda$(MeV) & E(MeV) & $R_{rms}$(fm) & $P_{S}(\%)$ & $P_{D1}(\%)$ & $P_{D2}(\%)$ & $P_{G1}(\%)$ & $P_{G2}(\%)$\tabularnewline
		\hline
		$\left[\Xi_{cc}^{*}\Xi_{cc}^{*}\right]_{J=0}^{I=1}$ & 2000 & 2.37 & 2.15 & 99.5 & 0.5 &  &  & \tabularnewline
		& 2200 & 7.46 & 1.45 & 98.9 & 1.1 &  &  & \tabularnewline
		& 2400 & 15.93 & 1.17 & 97.9 & 2.1 &  &  & \tabularnewline
		\hline
		$\left[\Xi_{cc}^{*}\Omega_{cc}^{*}\right]_{J=0}^{I=\frac{1}{2}}$ & 1800 & 1.11 & 2.75 & 99.8 & 0.2 &  &  & \tabularnewline
		& 2000 & 5.19 & 1.60 & 99.3 & 0.7 &  &  & \tabularnewline
		& 2200 & 12.93 & 1.22 & 98.5 & 1.5 &  &  & \tabularnewline
		\hline
		$\left[\Omega_{cc}^{*}\Omega_{cc}^{*}\right]_{J=0}^{I=0}$ & 1800 & 1.12 & 2.66 & 99.9 & 0.1 &  &  & \tabularnewline
		& 2000 & 5.50 & 1.50 & 99.6 & 0.4 &  &  & \tabularnewline
		& 2200 & 14.20 & 1.14 & 98.9 & 1.1 &  &  & \tabularnewline
		\hline
		$\left[\Xi_{cc}^{*}\Xi_{cc}^{*}\right]_{J=1}^{I=0}$ & 800 & 25.28 & 0.99 & 94.4 & 5.2 & 0.4 & 0.0 & \tabularnewline
		& 820 & 31.66 & 0.91 & 94.3 & 5.4 & 0.3 & 0.0 & \tabularnewline
		& 840 & 39.72 & 0.84 & 94.2 & 5.5 & 0.3 & 0.0 & \tabularnewline
		\hline
		$\left[\Xi_{cc}^{*}\Omega_{cc}^{*}\right]_{J=1}^{I=\frac{1}{2}}$ & 1800 & 1.07 & 2.90 & 97.9 & 2.1 & 0.0 & 0.0 & \tabularnewline
		& 2000 & 7.47 & 1.59 & 93.2 & 6.7 & 0.1 & 0.0 & \tabularnewline
		& 2200 & 20.89 & 1.25 & 88.6 & 11.1 & 0.3 & 0.0 & \tabularnewline
		\hline
		$\left[\Xi_{cc}^{*}\Xi_{cc}^{*}\right]_{J=2}^{I=1}$ & \multicolumn{8}{c}{$\times$}\tabularnewline
		\hline
		$\left[\Xi_{cc}^{*}\Omega_{cc}^{*}\right]_{J=2}^{I=\frac{1}{2}}$ & 2400 & 1.50 & 2.43 & 95.6 & 0.4 & 3.9 & 0.1 & \tabularnewline
		& 2500 & 5.04 & 1.62 & 91.9 & 0.7 & 7.2 & 0.2 & \tabularnewline
		& 2600 & 11.00 & 1.31 & 88.4 & 1.0 & 10.3 & 0.3 & \tabularnewline
		\hline
		$\left[\Omega_{cc}^{*}\Omega_{cc}^{*}\right]_{J=2}^{I=0}$ & 2400 & 2.71 & 1.92 & 95.8 & 0.4 & 3.7 & 0.1 & \tabularnewline
		& 2500 & 6.70 & 1.45 & 93.0 & 0.7 & 6.2 & 0.1 & \tabularnewline
		& 2600 & 13.03 & 1.21 & 90.0 & 0.9 & 8.9 & 0.2 & \tabularnewline
		\hline
		$\left[\Xi_{cc}^{*}\Xi_{cc}^{*}\right]_{J=3}^{I=0}$ & 1200 & 2.65 & 2.86 & 89.7 & 0.3 & 9.9 & 0.0 & 0.1\tabularnewline
		& 1300 & 5.29 & 2.34 & 88.5 & 0.3 & 11.0 & 0.0 & 0.2\tabularnewline
		& 1400 & 8.87 & 2.01 & 88.1 & 0.2 & 11.4 & 0.0 & 0.3\tabularnewline
		\hline
		$\left[\Xi_{cc}^{*}\Omega_{cc}^{*}\right]_{J=3}^{I=\frac{1}{2}}$ & 1120 & 8.39 & 1.02 & 99.7 & 0.0 & 0.3 & 0.0 & 0.0\tabularnewline
		& 1140 & 15.43 & 0.79 & 99.5 & 0.0 & 0.4 & 0.0 & 0.1\tabularnewline
		& 1160 & 25.78 & 0.64 & 99.2 & 0.0 & 0.7 & 0.0 & 0.1\tabularnewline
		\hline
	\end{tabular}
\end{table*}

\begin{figure*}
	\centering
	\subfigure[$V_{11}$]{
		\label{fig:vxx1011}
		\includegraphics[width=0.31\linewidth]{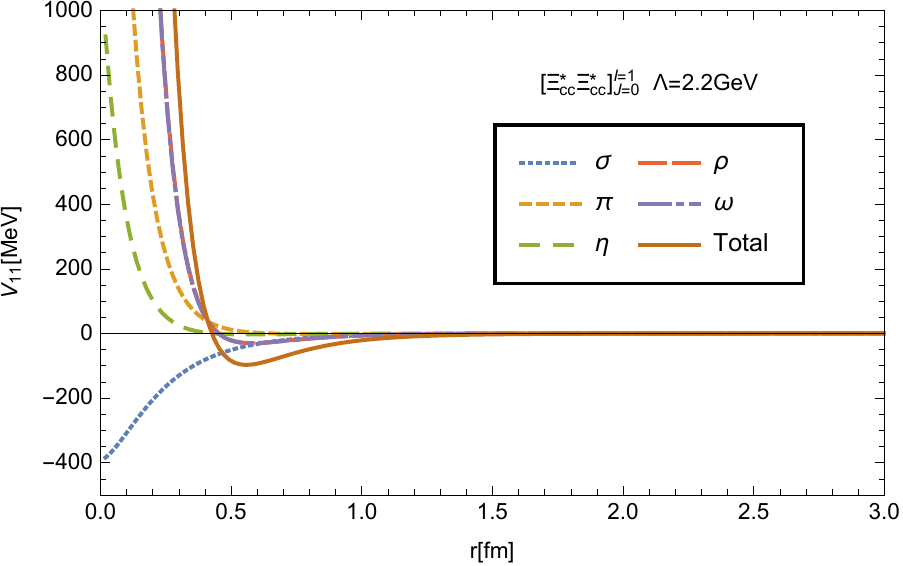}
	}
	\subfigure[$V_{12}$]{
		\label{fig:vxx1012}
		\includegraphics[width=0.31\linewidth]{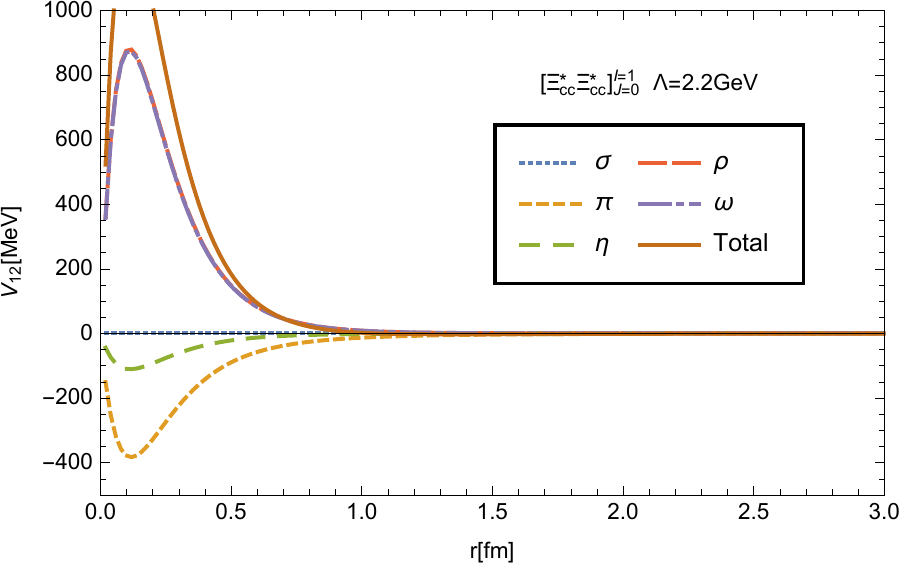}
	}
	\subfigure[$V_{22}$]{
		\label{fig:vxx1022}
		\includegraphics[width=0.31\linewidth]{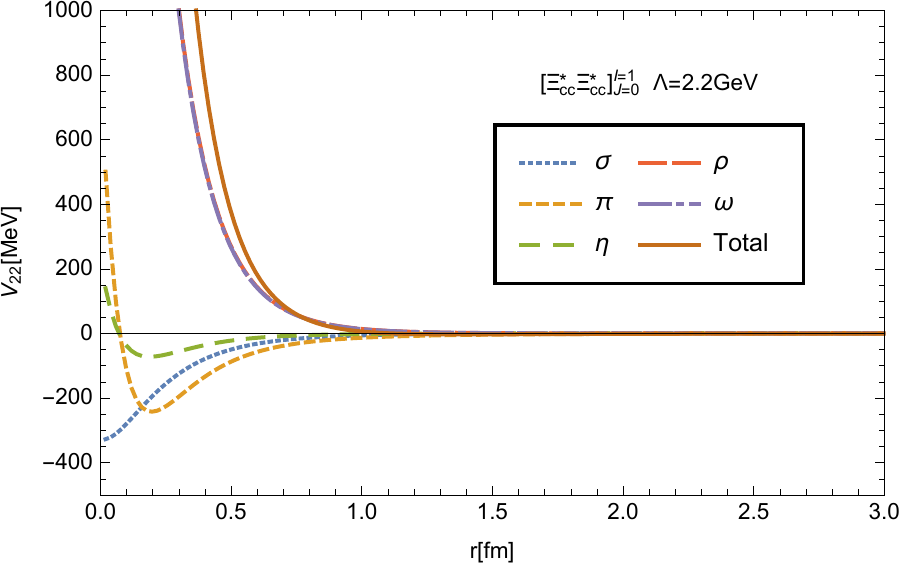}
	}
	\caption{The interactions potentials for the system $\left[\Xi_{cc}^*\Xi_{cc}^*\right]^{I=1}_{J=0}$. $V_{11}$, $V_{12}$ and $V_{22}$ denote the ${}^{1}S_{0}\leftrightarrow{}^{1}S_{0}$, ${}^{1}S_{0} \leftrightarrow{}^{5}D_{0}$ and ${}^{5}D_{0}\leftrightarrow{}^{5}D_{0}$ transitions potentials.}
	\label{potential:XX10}
\end{figure*}

For the $J=0$ systems, we consider the ${}^1S_0$ and ${}^5D_0$ wave
mixing effect. The contribution of the $D$-wave is actually quite
small, less than 2\%. For the
$\left[\Xi_{cc}^{*}\Xi_{cc}^{*}\right]_{J=0}^{I=1}$ system, we find
a loosely bound state with the binding energy 2.37-15.93 MeV, while
the cutoff parameter is 2.0-2.4 GeV. We present the potentials of
the system when the cutoff parameter is 2.2 GeV in
Fig.~\ref{potential:XX10}, where we use $V_{11}$, $V_{22}$ and
$V_{12}$ to denote the potentials for the $S$-wave channel, the
$D$-wave channel, and the off-diagonal term mixing the $S$-wave and
$D$-wave, respectively. The only attractive potential between the
baryons appears in the $S$-wave. For the
$\left[\Xi_{cc}^{*}\Omega_{cc}^{*}\right]_{J=0}^{I=\frac{1}{2}}$ and
$\left[\Omega_{cc}^{*}\Omega_{cc}^{*}\right]_{J=0}^{I=0}$ systems,
we get binding solutions when we change the cutoff parameter from
1.8 GeV to 2.2 GeV. Both of them have small binding energies and
reasonable root-mean-square radii.

For the $J=1$ systems, we calculate four channel mixing effects
among ${}^3S_1$, ${}^3D_1$, ${}^7D_1$ and ${}^7G_1$. Actually the
contribution of $G$-wave almost vanishes. For the
$\left[\Xi_{cc}^{*}\Xi_{cc}^{*}\right]_{J=1}^{I=0}$ system, a bound
state with the binding energy 25.28-39.72 MeV appears when the
cutoff parameter is chosen between 0.82-0.84 GeV. Given that the
binding energy is sensitive to the cutoff parameter, and the
root-mean-square radius is less than 1 fm, the system may not be a
perfect candidate of the molecular state. For the
$\left[\Xi_{cc}^{*}\Omega_{cc}^{*}\right]_{J=1}^{I=\frac{1}{2}}$
system, we obtain a bound state with binding energy 1.07-20.89 MeV
when the cutoff parameter is 1.8-2.2 GeV. The contributions of
$D$-wave channels increase with the cutoff parameter. The
contribution of the dominant $S$-wave channel is 88.6\% when the
cutoff is 2.2 GeV.

\begin{figure*}
	\centering
	\subfigure[$V_{11}$]{
		\label{fig:vxx1211}
		\includegraphics[width=0.31\linewidth]{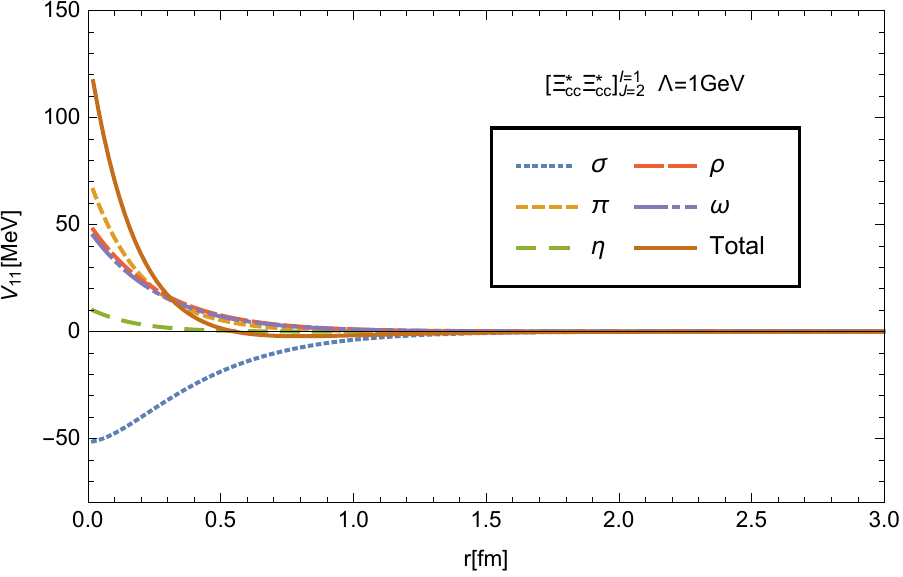}
	}
	\subfigure[$V_{22}$]{
		\label{fig:vxx1212}
		\includegraphics[width=0.31\linewidth]{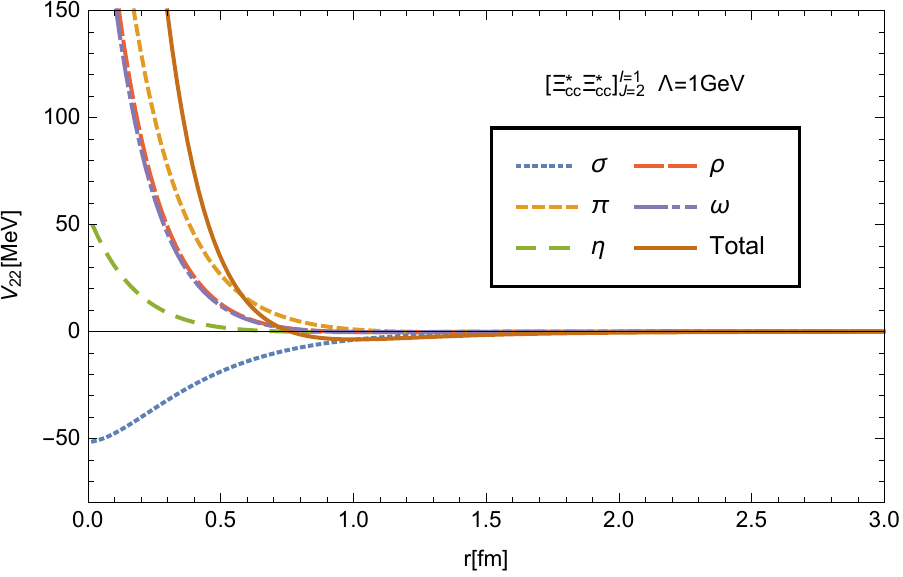}
	}
	\subfigure[$V_{33}$]{
		\label{fig:vxx1222}
		\includegraphics[width=0.31\linewidth]{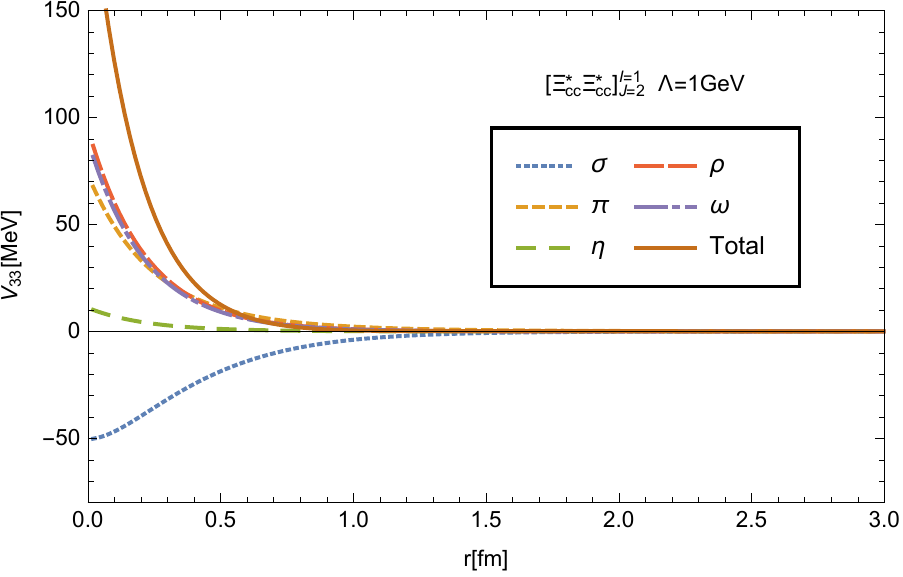}
	}
	\caption{The interactions potentials for the system $\left[\Xi_{cc}^*\Xi_{cc}^*\right]^{I=1}_{J=2}$. $V_{11}$, $V_{22}$ and $V_{33}$ denote the diagonal terms in the potential matrix for the channels ${}^{5}S_{2}$, ${}^{1}D_{2}$ and ${}^{5}D_{2}$ respectively.}
	\label{potential:XX12}
\end{figure*}

For the $\Xi_{cc}^{*}\Omega_{cc}^{*}$ and
$\Omega_{cc}^{*}\Omega_{cc}^{*}$ systems with total spin 2, we
obtain binding solutions with the cutoff parameter from 2.4 GeV to
2.6 GeV. The binding solution of the former system is 1.50-11.00
MeV, and 2.71-13.03 MeV for the latter one. There is no bound state
for the $\left[\Xi_{cc}^{*}\Xi_{cc}^{*}\right]_{J=2}^{I=1}$ system,
even if we include the couple-channel effect and vary the cutoff
parameter from 0.8 GeV to 3 GeV. The relevant interaction potentials
with the cutoff parameter 1.0 GeV is shown in
Fig.~\ref{potential:XX12}. The $S$-wave and $D$-wave potentials of
the system, $V_{11}$, $V_{22}$ and $V_{33}$, are hardly attractive.

For the $J=3$ case, there are only two possible systems,
$\left[\Xi_{cc}^{*}\Xi_{cc}^{*}\right]_{J=3}^{I=0}$ and
$\left[\Xi_{cc}^{*}\Omega_{cc}^{*}\right]_{J=3}^{I=\frac{1}{2}}$. We
consider the mixing effect of five channels, ${}^7S_3$, ${}^3D_3$,
${}^7D_3$, ${}^3G_3$ and ${}^7G_3$. The binding energy of the
$\left[\Xi_{cc}^{*}\Xi_{cc}^{*}\right]_{J=3}^{I=0}$ system is
2.65-8.87 MeV, when we change the cut off from 1.2 GeV to 1.4 GeV.
The contribution of the channel ${}^7D_3$ is important, the
percentage of which is over 10\%. For the
$\left[\Xi_{cc}^{*}\Omega_{cc}^{*}\right]_{J=3}^{I=\frac{1}{2}}$
system, the binding energy changes from 8.39 MeV to 25.78 MeV while
the cutoff parameter changes from 1.12 GeV to 1.16 GeV. In the wave
functions, the $S$-wave is dominant, whose contribution is over
99\%. The root-mean-square radius of the bound state is 0.64-1.02
fm, which seems a little small for a loosely bound state composed of
two doubly charmed baryons.

Among all possible systems composed of two spin-$\frac{3}{2}$
baryons, the systems
$\left[\Xi_{cc}^{*}\Xi_{cc}^{*}\right]_{J=0,3}^{I=1}$,
$\left[\Xi_{cc}^{*}\Omega_{cc}^{*}\right]_{J=0,1,2}^{I=\frac{1}{2}}$
and $\left[\Omega_{cc}^{*}\Omega_{cc}^{*}\right]_{J=0,2}^{I=0}$ are
good candidates of molecular states. For the system
$\left[\Xi_{cc}^{*}\Xi_{cc}^{*}\right]_{J=1}^{I=0}$ and
$\left[\Xi_{cc}^{*}\Omega_{cc}^{*}\right]_{J=3}^{I=\frac{1}{2}}$,
the existence of the bound solutions is very sensitive to cutoff
parameter. Meanwhile, their root-mean-square radii are less than
1fm. We do not find a binding solution for the
$\left[\Xi_{cc}^{*}\Xi_{cc}^{*}\right]_{J=2}^{I=1}$ system.

%%%%%%%%%%%%%%%%%%%%%%%%%%%%%%%%%%%%%%%%%%%%
\subsection{One spin-$\frac{3}{2}$ baryon and one spin-$\frac{1}{2}$ baryon system}\label{subsec_res2}
%%%%%%%%%%%%%%%%%%%%%%%%%%%%%%%%%%%%%%%%%%%%

We investigate the possible molecular systems composed of one
spin-$\frac{3}{2}$ baryon and one spin-$\frac{1}{2}$ baryon. Their
total spin can be 1 and 2. For the $J=1$ system, we consider the
couple-channel effect among ${}^3S_1$, ${}^3D_1$ and ${}^5D_1$
channels. For the $J=2$ system, the couple-channel effect is among
${}^5S_2$, ${}^3D_2$, ${}^5D_2$ and ${}^5G_2$ channels. We show the
binding energies, root-mean-square radii and contributions of
different channels of the $\Xi_{cc}^{*}\Xi_{cc}$ and
$\Omega_{cc}^{*}\Omega_{cc}$ systems in Table~\ref{Table_B3B1}.

\begin{table*}[htb]
	\centering
	\caption{The numerical results for the $B^*B$ systems. $\Lambda $ is the cutoff parameter. ``$B.E.$" is the binding energy. $R_{rms}$ is the root-mean-square radius. $P_{i}$ is the percentage of the different channels.}\label{Table_B3B1}
	
	\begin{tabular}{cccccccc}
		\hline
		States & $\Lambda$(MeV) & E(MeV) & $R_{rms}$(fm) & $P_{S}(\%)$ & $P_{D1}(\%)$ & $P_{D2}(\%)$ & $P_{F}(\%)$\tabularnewline
		\hline
		$\left[\Xi_{cc}^{*}\Xi_{cc}\right]_{J=1}^{I=0}$ & 1200 & 4.73 & 2.56 & 87.3 & 6.1 & 6.6 & \tabularnewline
		& 1400 & 8.29 & 2.15 & 88.0 & 5.4 & 6.6 & \tabularnewline
		& 1600 & 12.57 & 1.86 & 90.2 & 4.3 & 5.5 & \tabularnewline
		\hline
		$\left[\Xi_{cc}^{*}\Xi_{cc}\right]_{J=1}^{I=1}$ & 1000 & 1.85 & 2.17 & 98.9 & 0.6 & 0.5 & \tabularnewline
		& 1050 & 8.34 & 1.11 & 99.5 & 0.3 & 0.2 & \tabularnewline
		& 1100 & 24.39 & 0.67 & 99.9 & 0.1 & 0.0 & \tabularnewline
		\hline
		$\left[\Omega_{cc}^{*}\Omega_{cc}\right]_{J=1}^{I=0}$ & 1200 & 3.01 & 1.64 & 99.5 & 0.3 & 0.2 & \tabularnewline
		& 1250 & 8.97 & 1.03 & 99.6 & 0.2 & 0.2 & \tabularnewline
		& 1300 & 22.34 & 0.69 & 99.98 & 0.1 & 0.1 & \tabularnewline
		\hline
		$\left[\Omega_{cc}^{*}\Xi_{cc}\right]_{J=1}^{I=1/2}$ & 1200 & 4.36 & 1.48 & 99.3 & 0.6 & 0.1 & \tabularnewline
		& 1250 & 9.68 & 1.06 & 99.5 & 0.4 & 0.1 & \tabularnewline
		& 1300 & 18.74 & 0.79 & 99.8 & 0.2 & 0.0 & \tabularnewline
		\hline
		$\left[\Xi_{cc}^{*}\Omega_{cc}\right]_{J=1}^{I=1/2}$ & 1200 & 2.68 & 1.77 & 99.5 & 0.4 & 0.1 & \tabularnewline
		& 1250 & 6.81 & 1.19 & 99.6 & 0.3 & 0.1 & \tabularnewline
		& 1300 & 14.24 & 0.87 & 99.8 & 0.2 & 0.0 & \tabularnewline
		\hline
		$\left[\Xi_{cc}^{*}\Xi_{cc}\right]_{J=2}^{I=0}$ & 1200 & 2.33 & 2.89 & 91.4 & 0.5 & 8.1 & 0.0\tabularnewline
		& 1400 & 5.66 & 2.19 & 91.2 & 0.5 & 8.3 & 0.0\tabularnewline
		& 1600 & 10.15 & 1.80 & 92.6 & 0.4 & 7.0 & 0.0\tabularnewline
		\hline
		$\left[\Xi_{cc}^{*}\Xi_{cc}\right]_{J=2}^{I=1}$ & 1000 & 1.91 & 2.06 & 99.2 & 0.0 & 0.8 & 0.0\tabularnewline
		& 1050 & 8.78 & 1.05 & 99.6 & 0.0 & 0.4 & 0.0\tabularnewline
		& 1100 & 25.32 & 0.64 & 99.9 & 0.0 & 0.1 & 0.0\tabularnewline
		\hline
		$\left[\Omega_{cc}^{*}\Omega_{cc}\right]_{J=2}^{I=0}$ & 1200 & 2.97 & 1.62 & 99.6 & 0.0 & 0.4 & 0.0\tabularnewline
		& 1250 & 9.19 & 1.00 & 99.7 & 0.0 & 0.3 & 0.0\tabularnewline
		& 1300 & 23.06 & 0.67 & 99.9 & 0.0 & 0.1 & 0.0\tabularnewline
		\hline
		$\left[\Omega_{cc}^{*}\Xi_{cc}\right]_{J=2}^{I=1/2}$ & 1000 & 3.68 & 1.54 & 99.4 & 0.3 & 0.3 & 0.0\tabularnewline
		& 1050 & 16.00 & 0.86 & 99.5 & 0.2 & 0.3 & 0.0\tabularnewline
		& 1100 & 42.66 & 0.72 & 99.7 & 0.21 & 0.2 & 0.0\tabularnewline
		\hline
		$\left[\Xi_{cc}^{*}\Omega_{cc}\right]_{J=2}^{I=1/2}$ & 1000 & 1.53 & 2.17 & 99.6 & 0.2 & 0.2 & 0.0\tabularnewline
		& 1050 & 10.47 & 0.99 & 99.6 & 0.2 & 0.2 & 0.0\tabularnewline
		& 1100 & 32.40 & 0.64 & 99.7 & 0.2 & 0.1 & 0.0\tabularnewline
		\hline
	\end{tabular}
\end{table*}

For the $\left[\Xi_{cc}^{*}\Xi_{cc}\right]_{J=1}^{I=0}$ system, we
obtain a loosely bound state with binding energy 4.73-12.57 MeV
while the cutoff parameter is 1.2-1.6 GeV. The $D$-wave contribution
is about 10\% and decreases as the cutoff parameter increases. For
the $\left[\Xi_{cc}^{*}\Xi_{cc}\right]_{J=1}^{I=1}$ system, the
contribution of the $S$-wave is over 99\%. When we choose the cutoff
parameter at 1.05 GeV, the binding energy of the bound state is 8.34
MeV. For the $\left[\Omega_{cc}^{*}\Omega_{cc}\right]_{J=1}^{I=0}$
system, we find a bound state solution with binding energy 8.97
while the cutoff parameter is 1.25 GeV. For the
$\left[\Omega_{cc}^{*}\Xi_{cc}\right]_{J=1}^{I=1/2}$ system, a bound
solution with a dominant $S$-wave appears when the cutoff is 1.2
GeV-1.3 GeV. The result of
$\left[\Xi_{cc}^{*}\Omega_{cc}\right]_{J=1}^{I=1/2}$ is very similar
to that of $\left[\Omega_{cc}^{*}\Xi_{cc}\right]_{J=1}^{I=1/2}$. The
two systems are related by the $U$-spin and the $V$-spin symmetry.

For the systems with total spin 2, the contributions of the $G$-wave
channels are almost zero. For the
$\left[\Xi_{cc}^{*}\Xi_{cc}\right]_{J=2}^{I=0}$ system, a loosely
bound state with binding energy 2.33-10.15 MeV appears when the
cutoff parameter is 1.2-1.6 GeV. The contribution of ${}^5D_2$ is
about 8\% when the cutoff is 1.4 GeV. For the
$\left[\Xi_{cc}^{*}\Xi_{cc}\right]_{J=2}^{I=1}$ system, we obtain a
binding solution with the binding energy 8.78 MeV while the cutoff
parameter is 1.05 GeV. In the system, the $D$-waves contributions
are less than 1\%. For the
$\left[\Omega_{cc}^{*}\Omega_{cc}\right]_{J=2}^{I=0}$ system, we
obtain a binding solution with the binding energy 9.19 MeV when we
choose the cutoff parameter as 1.25 GeV. The $D$-waves contributions
are also very small. The results for
$\left[\Omega_{cc}^{*}\Xi_{cc}\right]_{J=2}^{I=1/2}$ and
$\left[\Xi_{cc}^{*}\Omega_{cc}\right]_{J=2}^{I=1/2}$ are almost the
same. When we choose the cutoff from 1.0 GeV to 1.1 GeV, we find a
bound state with binding energy 3.68-42.66 MeV for the former
system, and a binding solution with binding energy 1.53-32.4 MeV for
the latter one.

Considering the reasonable binding energies and
root-mean-square-radii of the above solutions, the
spin-$\frac{3}{2}$ baryon and spin-$\frac{1}{2}$ baryon systems are
all good candidates of molecular states.

%%%%%%%%%%%%%%%%%%%%%%%%%%%%%%%%%%%%%%%%%%%%
\subsection{The two doubly charmed systems with multi-channel mixing effect}\label{subsec_res3}
%%%%%%%%%%%%%%%%%%%%%%%%%%%%%%%%%%%%%%%%%%%%

We calculate the systems with channel mixing among $BB$, $B^*B$ and
$B^*B^*$ in this subsection. $B$ and $B^*$ are the
spin-$\frac{1}{2}$ and spin-$\frac{3}{2}$ doubly charmed baryons,
respectively. We present the possible systems with certain total
spin and isospin in Tables~\ref{Table_cccc1}-\ref{Table_cccc3}. In
the previous subsection, the couple-channel effect from the high
angular momentum states are small. The components of $G$-wave in the
total wave functions are usually negligible. Thus we only consider
the $D$-wave in our calculation.

\begin{table}[htb]
	\centering
	\caption{Possible mixing channels for the $\Omega_{cc}^{(*)}\Omega_{cc}^{(*)}$ system with $J=0$ and $J=2$.}\label{Table_cccc1}
	\begin{tabular}{c|c|ccc|ccc}
		\hline
		$I(J^{P})$ & $\Omega_{cc}\Omega_{cc}$ & \multicolumn{3}{c|}{$\Omega_{cc}^{*}\Omega_{cc}^{*}$} & \multicolumn{3}{c}{$\Omega_{cc}\Omega_{cc}^{*}$}\tabularnewline
		\hline
		$0(0^{+})$ & $|^{1}S_{0}\rangle$ & $|^{1}S_{0}\rangle$ & $|^{5}D_{0}\rangle$ &  & $|^{5}D_{0}\rangle$ &  & \tabularnewline
		$0(2^{+})$ & $|^{1}D_{2}\rangle$ & $|^{5}S_{2}\rangle$ & $|^{1}D_{2}\rangle$ & $|^{5}D_{2}\rangle$ & $|^{5}S_{2}\rangle$ & $|^{3}D_{2}\rangle$ & $|^{5}D_{2}\rangle$\tabularnewline
		\hline
	\end{tabular}
\end{table}

\begin{table*}[htb]
	\centering
	\caption{Possible mixing channels for the $\Xi_{cc}^{(*)}\Xi_{cc}^{(*)}$ system with total angular momentum 0, 1, 2 and 3.}\label{Table_cccc2}
	\begin{tabular}{c|cc|ccc|ccc}
		\hline
		$I(J^{P})$ & \multicolumn{2}{c|}{$\Xi_{cc}\Xi_{cc}$} & \multicolumn{3}{c|}{$\Xi_{cc}^{*}\Xi_{cc}^{*}$} & \multicolumn{3}{c}{$\Xi_{cc}\Xi_{cc}^{*}$}\tabularnewline
		\hline
		$0(1^{+})$ & $|^{3}S_{1}\rangle$ & $|^{3}D_{1}\rangle$ & $|^{3}S_{1}\rangle$ & $|^{3}D_{1}\rangle$ & $|^{7}D_{1}\rangle$ & $|^{3}S_{1}\rangle$ & $|^{3}D_{1}\rangle$ & $|^{5}D_{1}\rangle$\tabularnewline
		$0(3^{+})$ & $|^{3}D_{3}\rangle$ &  & $|^{7}S_{3}\rangle$ & $|^{3}D_{3}\rangle$ & $|^{7}D_{3}\rangle$ & $|^{3}D_{3}\rangle$ & $|^{5}D_{3}\rangle$ & \tabularnewline
		$1(0^{+})$ & $|^{1}S_{0}\rangle$ &  & $|^{1}S_{0}\rangle$ & $|^{5}D_{0}\rangle$ &  & $|^{5}D_{0}\rangle$ &  & \tabularnewline
		$1(2^{+})$ & $|^{1}D_{2}\rangle$ &  & $|^{5}S_{2}\rangle$ & $|^{1}D_{2}\rangle$ & $|^{5}D_{2}\rangle$ & $|^{5}S_{2}\rangle$ & $|^{3}D_{2}\rangle$ & $|^{5}D_{2}\rangle$\tabularnewline
		\hline
	\end{tabular}
\end{table*}

\begin{table*}[htb]
	\centering
	\caption{Possible mixing channels for the $\Xi_{cc}^{(*)}\Omega_{cc}^{(*)}$ system with total angular momentum 0, 1, 2 and 3.}\label{Table_cccc3}
	\begin{tabular}{c|cc|ccc|ccc|ccc}
		\hline
		$I(J^{P})$ & \multicolumn{2}{c|}{$\Xi_{cc}\Omega_{cc}$} & \multicolumn{3}{c|}{$\Xi_{cc}^{*}\Omega_{cc}^{*}$} & \multicolumn{3}{c|}{$\Xi_{cc}^{*}\Omega_{cc}$} & \multicolumn{3}{c}{$\Xi_{cc}\Omega_{cc}^{*}$}\tabularnewline
		\hline
		$\frac{1}{2}(0^{+})$ & $|^{1}S_{0}\rangle$ &  & $|^{1}S_{0}\rangle$ & $|^{5}D_{0}$ &  & $|^{5}D_{0}\rangle$ &  &  & $|^{5}D_{0}\rangle$ &  & \tabularnewline
		$\frac{1}{2}(1^{+})$ & $|^{3}S_{1}\rangle$ & $|^{3}D_{1}\rangle$ & $|^{3}S_{1}\rangle$ & $|^{3}D_{1}\rangle$ & $|^{7}D_{1}\rangle$ & $|^{3}S_{1}\rangle$ & $|^{3}D_{1}\rangle$ & $|^{5}D_{1}\rangle$ & $|^{3}S_{1}\rangle$ & $|^{3}D_{1}\rangle$ & $|^{5}D_{1}\rangle$\tabularnewline
		$\frac{1}{2}(2^{+})$ & $|^{1}D_{2}\rangle$ &  & $|^{5}S_{2}\rangle$ & $|^{1}D_{2}\rangle$ & $|^{5}D_{2}\rangle$ & $|^{5}S_{2}\rangle$ & $|^{3}D_{2}\rangle$ & $|^{5}D_{2}\rangle$ & $|^{5}S_{2}\rangle$ & $|^{3}D_{2}\rangle$ & $|^{5}D_{2}\rangle$\tabularnewline
		$\frac{1}{2}(3^{+})$ & $|^{3}D_{3}\rangle$ &  & $|^{7}S_{3}\rangle$ & $|^{3}D_{3}\rangle$ & $|^{7}D_{3}\rangle$ & $|^{3}D_{3}\rangle$ & $|^{5}D_{3}\rangle$ &  & $|^{3}D_{3}\rangle$ & $|^{5}D_{3}\rangle$ & \tabularnewline
		\hline
	\end{tabular}
\end{table*}

Because the masses of various systems are different, we define the
binding energy relative to the $BB$ threshold. When we solve the
coupled Schr\"odinger equations, we put the mass difference in the
kinetic term. The effective potentials as well as the spin and
orbital angular momentum dependent operators are consistent with
what we defined before. However, we ignore the spin-orbital coupling
effect in the off-diagonal elements of the interaction potentials.
Because the effect is dependent on the external momentum, we can
treat it as a high order correction compared with spin-spin and
tensor interactions. Thus, it is reasonable to use only spin-spin
and tensor interactions to describe the channel mixing effect. The
numerical results including binding energies, root-mean-square radii
and percentages of different channels are shown in
Tables~\ref{Table_bbbb1}-\ref{Table_bbbb3}.

\begin{table*}[htb]
	\centering
	\caption{The numerical results for the $\Omega^{(*)}\Omega^{(*)}$ systems. $\Lambda $ is the cutoff parameter. ``$B.E.$" is the binding energy. $R_{rms}$ is the root-mean-square radius.}\label{Table_bbbb1}
	\begin{tabular}{c|ccc|c|ccc|ccc}
		\hline
		$I(J^{P})$ & \multicolumn{3}{c|}{Result} & $\Omega_{cc}\Omega_{cc}$ & \multicolumn{3}{c|}{$\Omega_{cc}^{*}\Omega_{cc}^{*}$} & \multicolumn{3}{c}{$\Omega_{cc}\Omega_{cc}^{*}$}\tabularnewline
		\hline
		$0(0^{+})$ & $\Lambda$( MeV) & B.E.( MeV) & $R_{rms}$(fm) & $|^{1}S_{0}\rangle$ & $|^{1}S_{0}\rangle$ & $|^{5}D_{0}\rangle$ &  & $|^{5}D_{0}\rangle$ &  & \tabularnewline
		\cline{2-11}
		& 1300 & 5.64 & 1.34 & 94.8 & 4.1 & 0.1 &  & 1.0 &  & \tabularnewline
		& 1350 & 21.69 & 0.84 & 87.1 & 10.8 & 0.1 &  & 2.0 &  & \tabularnewline
		& 1400 & 57.81 & 0.61 & 78.9 & 18.5 & 0.0 &  & 2.6 &  & \tabularnewline
		\hline
		$0(2^{+})$ & $\Lambda$( MeV) & B.E.( MeV) & $R_{rms}$(fm) & $|^{1}D_{2}\rangle$ & $|^{5}S_{2}\rangle$ & $|^{1}D_{2}\rangle$ & $|^{5}D_{2}\rangle$ & $|^{5}S_{2}\rangle$ & $|^{3}D_{2}\rangle$ & $|^{5}D_{2}\rangle$\tabularnewline
		\cline{2-11}
		& 1360 & 10.24 & 0.50 & 0.6 & 6.5 & 0.1 & 0.4 & 92.4 & 0.0 & 0.0\tabularnewline
		& 1370 & 28.92 & 0.47 & 0.5 & 7.7 & 0.1 & 0.4 & 91.2 & 0.0 & 0.0\tabularnewline
		& 1380 & 49.90 & 0.45 & 0.5 & 8.9 & 0.1 & 0.4 & 90.1 & 0.0 & 0.0\tabularnewline
		\hline
	\end{tabular}
\end{table*}

\begin{table*}[htb]
	\centering
	\caption{The numerical results for the $\Xi^{(*)}\Xi^{(*)}$ systems. $\Lambda $ is the cutoff parameter. ``$B.E.$" is the binding energy. $R_{rms}$ is the root-mean-square radius.}\label{Table_bbbb2}
	\begin{tabular}{c|ccc|cc|ccc|ccc}
		\hline
		$I(J^{P})$ & \multicolumn{3}{c|}{Result} & \multicolumn{2}{c|}{$\Xi_{cc}\Xi_{cc}$} & \multicolumn{3}{c|}{$\Xi_{cc}^{*}\Xi_{cc}^{*}$} & \multicolumn{3}{c}{$\Xi_{cc}\Xi_{cc}^{*}$}\tabularnewline
		\hline
		$0(1^{+})$ & $\Lambda$( MeV) & B.E.( MeV) & $R_{rms}$(fm) & $|^{3}S_{1}\rangle$ & $|^{3}D_{1}\rangle$ & $|^{3}S_{1}\rangle$ & $|^{3}D_{1}\rangle$ & $|^{7}D_{1}\rangle$ & $|^{3}S_{1}\rangle$ & $|^{3}D_{1}\rangle$ & $|^{5}D_{1}\rangle$\tabularnewline
		\cline{2-12}
		& 800 & 36.47 & 1.01 & 52.9 & 0.0 & 31.9 & 0.0 & 0.2 & 14.4 & 0.1 & 0.5\tabularnewline
		& 810 & 42.61 & 0.96 & 50.8 & 0.0 & 33.9 & 0.0 & 0.2 & 14.5 & 0.1 & 0.5\tabularnewline
		& 820 & 49.34 & 0.92 & 49.0 & 0.0 & 35.9 & 0.0 & 0.1 & 14.5 & 0.0 & 0.5\tabularnewline
		\hline
		$0(3^{+})$ & $\Lambda$( MeV) & B.E.( MeV) & $R_{rms}$(fm) & $|^{3}D_{3}\rangle$ &  & $|^{7}S_{3}\rangle$ & $|^{3}D_{3}\rangle$ & $|^{7}D_{3}\rangle$ & $|^{3}D_{3}\rangle$ & $|^{5}D_{3}\rangle$ & \tabularnewline
		\cline{2-12}
		& 1370 & 5.50 & 1.35 & 0.4 &  & 58.4 & 0.8 & 0.9 & 11.3 & 28.2 & \tabularnewline
		& 1380 & 11.01 & 1.32 & 0.4 &  & 58.2 & 0.9 & 0.9 & 11.3 & 28.4 & \tabularnewline
		& 1400 & 22.69 & 1.23 & 0.5 &  & 57.4 & 1.2 & 1.1 & 11.2 & 28.6 & \tabularnewline
		\hline
		$1(0^{+})$ & $\Lambda$( MeV) & B.E.( MeV) & $R_{rms}$(fm) & $|^{1}S_{0}\rangle$ &  & $|^{1}S_{0}\rangle$ & $|^{5}D_{0}\rangle$ &  & $|^{5}D_{0}\rangle$ &  & \tabularnewline
		\cline{2-12}
		& 1060 & 0.87 & 2.91 & 95.9 &  & 3.0 & 0.1 &  & 1.0 &  & \tabularnewline
		& 1100 & 8.83 & 1.21 & 86.4 &  & 10.8 & 0.1 &  & 2.7 &  & \tabularnewline
		& 1150 & 35.62 & 0.78 & 76.5 &  & 19.7 & 0.1 &  & 3.7 &  & \tabularnewline
		\hline
		$1(2^{+})$ & $\Lambda$( MeV) & B.E.( MeV) & $R_{rms}$(fm) & $|^{1}D_{2}\rangle$ &  & $|^{5}S_{2}\rangle$ & $|^{1}D_{2}\rangle$ & $|^{5}D_{2}\rangle$ & $|^{5}S_{2}\rangle$ & $|^{3}D_{2}\rangle$ & $|^{5}D_{2}\rangle$\tabularnewline
		\cline{2-12}
		& 1120 & 13.26 & 0.58 & 0.7 &  & 6.7 & 0.2 & 0.4 & 92.0 & 0.0 & 0.0\tabularnewline
		& 1130 & 29.16 & 0.54 & 0.6 &  & 8.0 & 0.2 & 0.4 & 90.8 & 0.0 & 0.0\tabularnewline
		& 1140 & 47.10 & 0.51 & 0.5 &  & 9.4 & 0.2 & 0.4 & 89.5 & 0.0 & 0.0\tabularnewline
		\hline
	\end{tabular}
\end{table*}

\begin{table*}[htb]
	\centering
	\caption{The numerical results for the $\Xi^{(*)}\Omega^{(*)}$ systems. $\Lambda $ is the cutoff parameter. ``$B.E.$" is the binding energy. $R_{rms}$ is the root-mean-square radius.}\label{Table_bbbb3}
	\begin{tabular}{c|ccc|cc|ccc|ccc|ccc}
		\hline
		$I(J^{P})$ & \multicolumn{3}{c|}{Result} & \multicolumn{2}{c|}{$\Xi_{cc}\Omega_{cc}$} & \multicolumn{3}{c|}{$\Xi_{cc}^{*}\Omega_{cc}^{*}$} & \multicolumn{3}{c|}{$\Xi_{cc}^{*}\Omega_{cc}$} & \multicolumn{3}{c}{$\Xi_{cc}\Omega_{cc}^{*}$}\tabularnewline
		\hline
		$\frac{1}{2}(0^{+})$ & $\Lambda$(MeV) & B.E.(MeV) & $R_{rms}$(fm) & $|^{1}S_{0}\rangle$ &  & $|^{1}S_{0}\rangle$ & $|^{5}D_{0}\rangle$ &  & $|^{5}D_{0}\rangle$ &  &  & $|^{5}D_{0}\rangle$ &  & \tabularnewline
		\cline{2-15}
		& 1170 & 1.21 & 2.50 & 97.0 &  & 2.3 & 0.0 &  & 0.3 &  &  & 0.3 &  & \tabularnewline
		& 1200 & 5.98 & 1.35 & 92.0 &  & 6.5 & 0.0 &  & 0.8 &  &  & 0.7 &  & \tabularnewline
		& 1250 & 28.64 & 0.81 & 81.7 &  & 15.7 & 0.0 &  & 1.3 &  &  & 1.3 &  & \tabularnewline
		\hline
		$\frac{1}{2}(1^{+})$ & $\Lambda$(MeV) & B.E.(MeV) & $R_{rms}$(fm) & $|^{3}S_{1}\rangle$ & $|^{3}D_{1}\rangle$ & $|^{3}S_{1}\rangle$ & $|^{3}D_{1}\rangle$ & $|^{7}D_{1}\rangle$ & $|^{3}S_{1}\rangle$ & $|^{3}D_{1}\rangle$ & $|^{5}D_{1}\rangle$ & $|^{3}S_{1}\rangle$ & $|^{3}D_{1}\rangle$ & $|^{5}D_{1}\rangle$\tabularnewline
		\cline{2-15}
		& 1200 & 5.97 & 1.38 & 93.4 & 0.0 & 5.1 & 0.0 & 0.1 & 0.3 & 0.1 & 0.4 & 0.2 & 0.1 & 0.4\tabularnewline
		& 1220 & 12.05 & 1.10 & 90.2 & 0.0 & 7.8 & 0.0 & 0.0 & 0.4 & 0.2 & 0.5 & 0.2 & 0.2 & 0.5\tabularnewline
		& 1240 & 21.11 & 0.94 & 87.0 & 0.0 & 10.7 & 0.0 & 0.0 & 0.4 & 0.3 & 0.6 & 0.3 & 0.2 & 0.5\tabularnewline
		\hline
		$\frac{1}{2}(2^{+})$ & $\Lambda$(MeV) & B.E.(MeV) & $R_{rms}$(fm) & $|^{1}D_{2}\rangle$ &  & $|^{5}S_{2}\rangle$ & $|^{1}D_{2}\rangle$ & $|^{5}D_{2}\rangle$ & $|^{5}S_{2}\rangle$ & $|^{3}D_{2}\rangle$ & $|^{5}D_{2}\rangle$ & $|^{5}S_{2}\rangle$ & $|^{3}D_{2}\rangle$ & $|^{5}D_{2}\rangle$\tabularnewline
		\cline{2-15}
		& 1080 & 2.24 & 0.67 & 0.0 &  & 0.0 & 0.0 & 0.0 & 59.0 & 0.2 & 0.0 & 40.7 & 0.1 & 0.0\tabularnewline
		& 1100 & 29.0 & 0.59 & 0.0 &  & 0.0 & 0.0 & 0.0 & 56.2 & 0.2 & 0.0 & 43.5 & 0.1 & 0.0\tabularnewline
		& 1120 & 61.28 & 0.54 & 0.0 &  & 0.0 & 0.0 & 0.0 & 54.3 & 0.1 & 0.0 & 45.5 & 0.1 & 0.0\tabularnewline
		\hline
		$\frac{1}{2}(3^{+})$ & $\Lambda$(MeV) & B.E.(MeV) & $R_{rms}$(fm) & $|^{3}D_{3}\rangle$ &  & $|^{7}S_{3}\rangle$ & $|^{3}D_{3}\rangle$ & $|^{7}D_{3}\rangle$ & $|^{3}D_{3}\rangle$ & $|^{5}D_{3}\rangle$ &  & $|^{3}D_{3}\rangle$ & $|^{5}D_{3}\rangle$ & \tabularnewline
		\cline{2-15}
		& 1290 & 7.28 & 0.55 & 0.0 &  & 93.76 & 0.0 & 0.0 & 0.4 & 2.8 &  & 0.4 & 2.8 & \tabularnewline
		& 1300 & 23.69 & 0.53 & 0.0 &  & 93.97 & 0.0 & 0.0 & 0.4 & 2.7 &  & 0.4 & 2.7 & \tabularnewline
		& 1310 & 41.13 & 0.51 & 0.1 &  & 93.8 & 0.0 & 0.0 & 0.4 & 2.6 &  & 0.4 & 2.7 & \tabularnewline
		\hline
	\end{tabular}
\end{table*}

For the $\Omega_{cc}^{(*)}\Omega_{cc}^{(*)}$ system with
$I(J^P)=0(0^+)$, we find a binding solution when the cutoff is
around 1.3 GeV. The dominant component of this solution is
$\Omega_{cc}\Omega_{cc}|^1S_0\rangle$. The component of
$\Omega_{cc}^*\Omega_{cc}^*|^1S_0\rangle$ increases with the cutoff
parameter. The $D$-wave contributions to the system are
insignificant. For the system with $I(J^P)=0(2^+)$, we find a
solution with the main component
$\Omega_{cc}\Omega_{cc}^*|^5S_2\rangle$, which agrees with the
single channel results in subsection~\ref{subsec_res2}. The mixing
effect of the channel $\Omega_{cc}^*\Omega_{cc}^*|^5S_2\rangle$ is
about 10\% and should not be ignored.

For the $\Xi_{cc}^{(*)}\Xi_{cc}^{(*)}$ system with $I(J^P)=0(1^+)$,
we find a binding solution with a small cutoff, 0.8 GeV. The channel
mixing effect is very prominent among the $\Xi_{cc}\Xi_{cc}$,
$\Xi_{cc}^{*}\Xi_{cc}^{*}$ and $\Xi_{cc}\Xi_{cc}^{*}$ with $S$-wave.
Their contributions to the total wave function are about 50\%, 35\%
and 15\%, respectively. The $D$-wave channels mixing effect in this
system is tiny. For the $I(J^P)=0(3^+)$ system, the channel mixing
effect is also important. We obtain a binding solution with cutoff
around 1.38 GeV. In the system, it is interesting that the $D$-wave
contribution of $\Xi_{cc}\Xi_{cc}^{*}$ reaches up to 28\%. For the
$I(J^P)=1(0^+)$ system, a binding solution with the main channel
$\Xi_{cc}\Xi_{cc}|^1S_0\rangle$ appears when the cutoff is 1.06 GeV.
For the $I(J^P)=1(2^+)$ system, we obtain a binding solution based
on $\Xi_{cc}\Xi_{cc}^*|^3S_1\rangle$. The result is consistent with
the calculation considering only the couple-channel effect inside
the $\Xi_{cc}\Xi_{cc}^{*}$ systems. The $\Xi_{cc}^{*}\Xi_{cc}^{*}$
channels  contribute 7\%-10\% to the wave function, which makes the
cutoff parameter a little larger when the binding energy is the
same.

The isospin of the $\Xi_{cc}^{(*)}\Omega_{cc}^{(*)}$ systems is
$\frac{1}{2}$. For the system with $J=0$, we find a bound state with
binding energy 1.21-28.64 MeV. The 97\% component of the system is
the $\Xi_{cc}\Omega_{cc}|^3S_0\rangle$ state when the cutoff is 1.17
GeV. The main part of the rest components is the $S$-wave of
$\Xi_{cc}^*\Omega_{cc}^*$. For the $I(J^P)=\frac{1}{2}(1^+)$ system,
we obtain a binding solution when the cutoff is around 1.2 GeV. The
system has a main part of the $\Xi_{cc}\Omega_{cc}|^1S_0\rangle$
channel, which mixes with the channel
$\Xi_{cc}^*\Omega_{cc}^*|^1S_0\rangle$. For the system with $J=2$, a
bound state solution appears when the cutoff is 1.1 GeV. The system
is dominated by the $S$-waves of $\Xi_{cc}^*\Omega_{cc}$ and
$\Xi_{cc}\Omega_{cc}^*$, and their contributions are 56.2\% and
43.5\%, respectively. For the system with $I(J^P)=\frac{1}{2}(3^+)$,
we obtain a binding solution when the cutoff is around 1.3 GeV. In
this system, the contribution of the
$\Xi_{cc}^*\Omega_{cc}^*|^7S_3\rangle$ channel is over 93\%, which
is the only possible $S$-wave channel. The main component of the
other channels are the $|^5D_3\rangle$ states of
$\Xi_{cc}^*\Omega_{cc}$ and $\Xi_{cc}\Omega_{cc}^*$, each of which
has the contribution about 2.7\%.

\begin{figure*}
	\centering
	\subfigure[$V_{11}$]{
		\label{fig:v00+1300-p11}
		\includegraphics[width=0.31\linewidth]{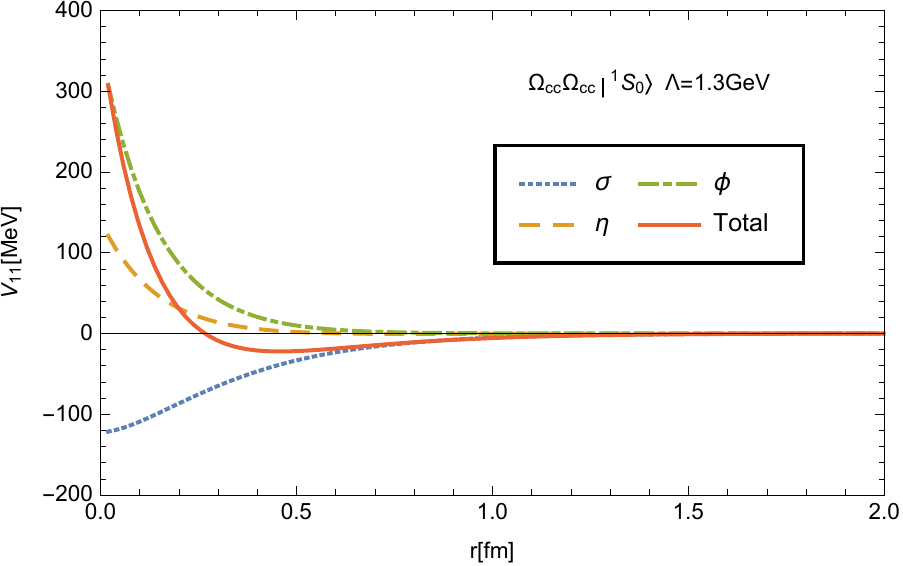}
	}
	\subfigure[$V_{22}$]{
		\label{fig:v00+1300-p22}
		\includegraphics[width=0.31\linewidth]{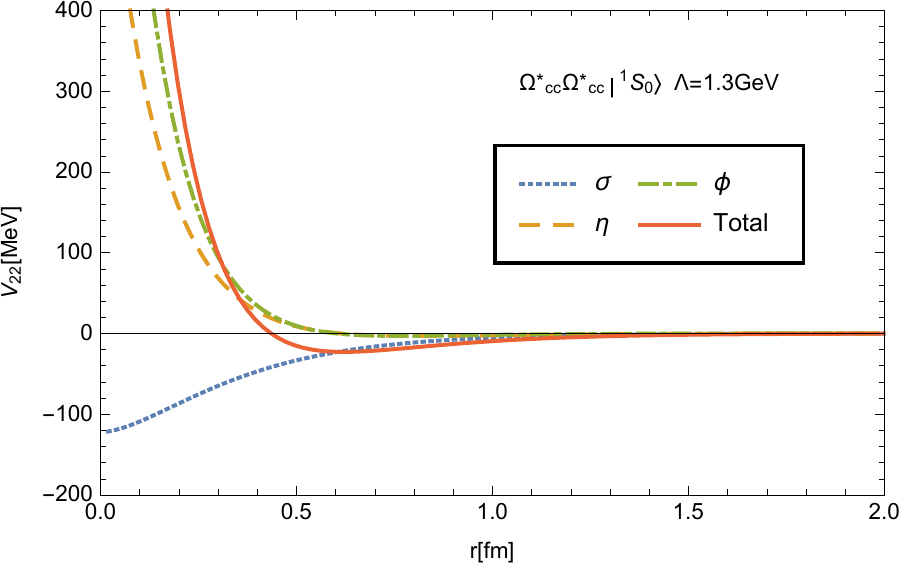}
	}
	\subfigure[$V_{44}$]{
		\label{fig:v00+1300-p44}
		\includegraphics[width=0.31\linewidth]{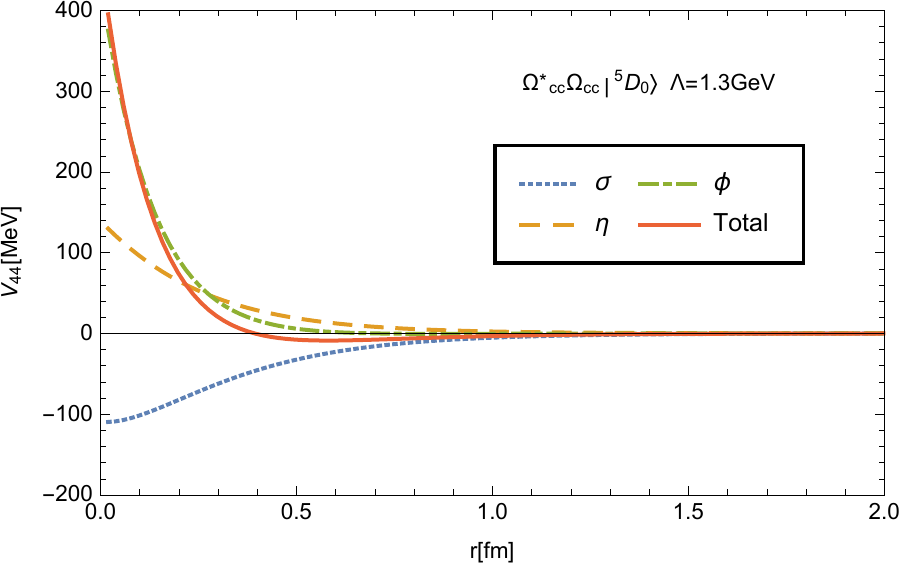}
	}
	\caption{The interactions potentials for the system with$I(J^P)=0(0^+)$. $V_{11}$, $V_{22}$ and $V_{44}$ denote the diagonal terms in the potential matrix for the channels $\Omega_{cc}\Omega_{cc}{|}^{1}S_{0}\rangle$, $\Omega_{cc}^*\Omega_{cc}^*{|}^{1}S_{0}\rangle$ and $\Omega_{cc}^*\Omega_{cc}{|}^{5}D_{0}\rangle$ respectively.}
	\label{potential:00+}
\end{figure*}

In Ref.~\cite{Meng:2017fwb}, the authors studied the systems composed of two spin-$\frac{1}{2}$ doubly charmed baryons.
We show the results in Table~\ref{Table_compare} and make some comparison.
After considering the channel mixing effect among the $BB$, $B^*B$ and $B^*B^*$ systems, we get some different conclusions for the $0(0^+)$, $1(0^+)$ and $\frac{1}{2}(0^+)$ systems.
As an example, we show the potentials of the $0(0^+)$ systems in Fig.~\ref{potential:00+}, where we give the potentials of the main channels.
In the single channel calculation Ref.~\cite{Meng:2017fwb}, the attraction of the $\Omega_{cc}\Omega_{cc}$ system is too weak to produce a bound state.
But with the help of the channel mixing effects, especially the $\Omega_{cc}^*\Omega_{cc}^*|^1S_0\rangle$ channel, we can obtain a binding solution.
The channel mixing effect plays the same role for the $1(0^+)$ and $\frac{1}{2}(0^+)$ systems.

Most of the $B^{(*)}B^{(*)}$ systems with the channel mixing effect are dominated by one channel with the percentage about 90\%.
However, the couple-channel effect from other channels may produce a bound state, which does not exist in the single channel case.
Another interesting observation is that the two or three main components of the systems are comparable, which may cause a large shift of the binding energy.
Actually the numerical results are quite complicated with the couple-channel effect.
The binding solutions become even more sensitive to the cutoff parameter.

\begin{table*}
	\centering
	\caption{The comparison of the $BB$ systems with and without channel mixing effect for the systems $B^*B^*$ and $B^*B$.}\label{Table_compare}
	\begin{tabular}{c|ccc|ccc}
		\hline
		& \multicolumn{3}{c|}{This work} & \multicolumn{3}{c}{\cite{Meng:2017fwb}}\tabularnewline
		\hline
		$I(J^{P})$ & $\Lambda$(MeV) & B.E.(MeV) & $R_{rms}$(fm) & $\Lambda$(MeV) & B.E.(MeV) & $R_{rms}$(fm)\tabularnewline
		\hline
		$\frac{1}{2}(1^{+})$ & 1200 & 5.97 & 1.38 & 1200 & 0.56 & 3.45\tabularnewline
		$0(1^{+})$ & 800 & 36.47 & 1.01 & 1100 & 0.68 & 3.32\tabularnewline
		$0(0^{+})$ & 1300 & 5.64 & 1.34 & \multicolumn{3}{c}{$\times$}\tabularnewline
		$1(0^{+})$ & 1060 & 0.87 & 2.91 & \multicolumn{3}{c}{$\times$}\tabularnewline
		$\frac{1}{2}(0^{+})$ & 1170 & 1.21 & 2.50 & \multicolumn{3}{c}{$\times$}\tabularnewline
		\hline
	\end{tabular}
\end{table*}

%%%%%%%%%%%%%%%%%%%%%%%%%%%%%%%%%%%%%%%%%%%%
\section{Discussions and conclusions}\label{sec_dis}
%%%%%%%%%%%%%%%%%%%%%%%%%%%%%%%%%%%%%%%%%%%%

In this work, with the help of the one-boson-exchange model, we
systematically investigate the possible molecular systems composed
of two spin-$\frac{3}{2}$ doubly charmed baryons, $B^{*}B^{*}$, as
well as the systems composed of one spin-$\frac{3}{2}$ and one
spin-$\frac{1}{2}$ doubly charmed baryon, $B^{*}B$.
We also study the couple-channel effect between various combinations of baryons.
We consider the $S$-waves and $D$-waves of the $BB$, $B^*B^*$ and $BB^*$ systems together.
The binding energies are defined relative to the $BB$ threshold, and the mass differences of different channels are put in the kinetic terms when calculating the coupled Schr\"odinger equations.

For the two spin-$\frac{3}{2}$ doubly charmed baryons systems, we
consider the channels mixing among possible $S$-wave $D$-wave and
$G$-wave. After considering the binding energies and
root-mean-square radii, as well as the reasonable value of cutoff
parameter, the following systems are good candidate of molecular
states, $\left[\Xi_{cc}^{*}\Xi_{cc}^{*}\right]_{J=0,3}^{I=1}$,
$\left[\Xi_{cc}^{*}\Omega_{cc}^{*}\right]_{J=0,1,2}^{I=\frac{1}{2}}$
and $\left[\Omega_{cc}^{*}\Omega_{cc}^{*}\right]_{J=0,2}^{I=0}$.

For the system composed of one spin-$\frac{3}{2}$ baryon and one
spin-$\frac{1}{2}$ baryon, their total spin can be 1 or 2. We
consider the channels mixing among ${}^3S_1$, ${}^3D_1$, ${}^5D_1$
for the $J=1$ system. For the system with $J=2$, we consider the
channels mixing between ${}^5S_2$, ${}^3D_2$, ${}^5D_2$ and
${}^5G_2$. The spin-$\frac{3}{2}$ baryon and spin-$\frac{1}{2}$
baryon systems are all candidates of molecular states, such as
$\left[\Xi_{cc}^{*}\Xi_{cc}\right]_{J=1,2}^{I=0,1}$,
$\left[\Omega_{cc}^{*}\Omega_{cc}\right]_{J=1,2}^{I=0}$ and
$\left[\Xi_{cc}^{*}\Omega_{cc}(\Omega_{cc}^{*}\Xi_{cc})\right]_{J=1,2}^{I=\frac{1}{2}}$.
The systems $\left[\Xi_{cc}^{*}\Xi_{cc}\right]_{J=1,2}^{I=0}$ are
particularly interesting. Both of them have small binding energies
around several MeVs, and large root-mean-square radii, 2-3 fm, when
the cutoff is from 1.2 GeV to 1.6 GeV.

We also consider the channel mixing between the $S$-wave and
$D$-wave $BB$, $B^*B^*$ and $BB^*$ states. Most of the
$B^{(*)}B^{(*)}$ systems are dominated by one single channel, which
has about 90\% contribution. However, the channel mixing effect from
the other channels may produce a bound state, which does not exist
in the single channel case, such as the systems with $J=0$.
Moreover, the two or three main components of the systems are
sometimes comparable, which may cause a large shift of the binding
energy, such as the system with $I(J^P)=0(1^+)$.

As a byproduct, we consider the systems composed of one baryon and
one antibaryon, $B^{*}\bar{B}^{*}$ and $B^{*}\bar{B}$. The results
are collected in Appendix~\ref{app_BBbar}. The baryon-antibaryon
systems may not be stable, because of the three meson decay modes
through quark rearrangement when the threshold is open. Although
molecular states composed of two doubly charmed baryons seem too
hard to be produced in the experiment, some possible structure
composed of one baryon and one antibaryon may be discovered at LHCb
in the future if they are not very broad.

%%%%%%%%%%%%%%%%%%%
\bigskip\noindent\textbf{Acknowledgements:}
B. Yang is very grateful to X.Z Weng and G.J. Wang for very
helpful discussions. This project is supported by the National
Natural Science Foundation of China under Grants 11575008,
11621131001 and National Key Basic Research Program of China
(2015CB856700).

\begin{appendix}
	
\section{Fourier transformation formulae}\label{app_Four}

The specific expressions of the scalar functions $H_i=H_i(\Lambda,m_{\sigma/\alpha/\beta},r)$ and $M_i=M_i(\Lambda,m_{\alpha},r)$ are as follows,
\begin{eqnarray}
H_{0}(\Lambda,m,r)&=Y(ur)-\frac{\lambda}{u}Y(\lambda r)-\frac{r\beta^{2}}{2u}Y(\lambda r),
\nonumber\\
H_{1}(\Lambda,m,r)&=Y(ur)-\frac{\lambda}{u}Y(\lambda r)-\frac{r\lambda^{2}\beta^{2}}{2u^{3}}Y(\lambda r),
\nonumber\\
H_{2}(\Lambda,m,r)&=Z_{1}(ur)-\frac{\lambda^{3}}{u^{3}}Z_{1}(\lambda r)-\frac{\lambda\beta^{2}}{2u^{3}}Y(\lambda r),
\nonumber\\
H_{3}(\Lambda,m,r)&=Z(ur)-\frac{\lambda^{3}}{u^{3}}Z(\lambda r)-\frac{\lambda\beta^{2}}{2u^{3}}Z_{2}(\lambda r),
\nonumber\\
M_{0}(\Lambda,m,r)&=-\frac{1}{\theta r}\left[\cos(\theta r)-e^{-\lambda r}\right]+\frac{\beta^{2}}{2\theta\lambda}e^{-\lambda r},
\nonumber\\
M_{1}(\Lambda,m,r)&=-\frac{1}{\theta r}\left[\cos(\theta r)-e^{-\lambda r}\right]-\frac{\lambda\beta^{2}}{2\theta^{3}}e^{-\lambda r},
\nonumber\\
M_{3}(\Lambda,m,r)&=-\left[\cos(\theta r)-\frac{3\sin(\theta r)}{\theta r}-3\frac{\cos(\theta r)}{\theta^{2}r^{2}}\right]\frac{1}{\theta r}
\nonumber\\
&-\frac{\lambda^{3}}{\theta^{3}}Z(\lambda r)-\frac{\lambda\beta^{2}}{2\theta^{3}}Z_{2}(\lambda r),
\end{eqnarray}
where
\[\beta^{2}=\Lambda^{2}-m^{2},~~u^{2}=m^{2}-Q_{0}^{2},~~\]
\[\theta^{2}=-(m^{2}-Q_{0}^{2}),~~\lambda^{2}=\Lambda^{2}-Q_{0}^{2},\]
and
\[Y(x)=\frac{e^{-x}}{x},~~Z(x)=(1+\frac{3}{x}+\frac{3}{x^{2}})Y(x),\]
\[Z_{1}(x)=(\frac{1}{x}+\frac{1}{x^{2}})Y(x),~~Z_{2}(x)=(1+x)Y(x).\]
$Q_0$ in the expression is the zeroth component of the four momentum of exchanged meson.

The above scalar functions come from Fourier transformation.
We give some useful Fourier transformation formulae.
\begin{eqnarray}
\frac{1}{u^{2}+\bm{Q}^{2}}\mathcal{F}^{2}(Q)&\rightarrow\frac{u}{4\pi}H_{0}(\Lambda,m,r),
\nonumber\\
\frac{\bm{Q}^{2}}{u^{2}+\bm{Q}^{2}}\mathcal{F}^{2}(Q)&\rightarrow-\frac{u^{3}}{4\pi}H_{1}(\Lambda,m,r),
\nonumber\\
\frac{\bm{Q}}{u^{2}+\bm{Q}^{2}}\mathcal{F}^{2}(Q)&\rightarrow\frac{iu^{3}}{4\pi}\boldsymbol{r}H_{2}(\Lambda,m,r),
\nonumber\\
\frac{Q_{i}Q_{j}}{u^{2}+\bm{Q}^{2}}\mathcal{F}^{2}(Q)&\rightarrow-\frac{u^{3}}{12\pi}[H_{3}(\Lambda,m,r)K_{ij}
\nonumber\\
&~~~~+H_{1}(\Lambda,m,r)\delta_{ij}].
\end{eqnarray}
If $u_{ex}^2=m_{ex}^2-Q_0^2<0$, the last formula above changes into
\begin{equation}
\begin{split}
\frac{Q_{i}Q_{j}}{u^{2}+\bm{Q}^{2}}\mathcal{F}^{2}(Q)\rightarrow
-\frac{\theta^{3}}{12\pi}&[M_{3}(\Lambda,m,r)K_{ij}  \\
&+M_{1}(\Lambda,m,r)\delta_{ij}].
\end{split}
\end{equation}

\section{Operators}\label{app_ope}

We extract some specific structures in the effective potentials and express them as some angular momentum dependent operators, $\Delta_{SS}$, $\Delta_{LS}$ and $\Delta_{T}$.
For the system with spin-$\frac{1}{2}$ initial and final states, their structures are as follows,
\begin{eqnarray}
&\Delta_{S_AS_B}=\bm{\sigma}\cdot\bm{\sigma},~\Delta_{LS}=\bm{L}\cdot\bm{S},\nonumber \\ &\Delta_{T}={3\bm{\sigma}\cdot\bm{r}\bm{\sigma}\cdot\bm{r}\over r^2}-\bm{\sigma}\cdot\bm{\sigma},
\end{eqnarray}
where $\bm{L}$ and $\bm{S}$ are the orbital angular momentum and total spin of two baryons.
$\bm{\sigma}$ is the Pauli matrix.

For a spin-$\frac{3}{2}$ baryon, we introduce the Rarita-Schwinger field $\Psi^{\mu}$.
The field is defined through
\begin{equation}
\Psi^\mu(\lambda)=\underset{m_\lambda}{\sum}\underset{m_s}{\sum}\braket{
	1m_\lambda,{1\over 2} m_s|{3\over 2}}\epsilon^\mu(m_\lambda)\chi(m_s)=S^\mu_t\Phi(\lambda),
\end{equation}
where $\chi(m_s)$ is a two-component spinor with the third component of spin $m_s$.
$\epsilon^\mu(m_\lambda)$ is the polarization vector of a $J=1$ field with the third component of spin $m_{\lambda}$,
\begin{equation}
\begin{split}
\epsilon^\mu(+)&=-\frac{1}{\sqrt{2}}\left[0,1,i,0\right]^T,~
\epsilon^\mu(0)=\left[0,0,0,1\right]^T,
\\
\epsilon^\mu(-)&=\frac{1}{\sqrt{2}}\left[0,1,-i,0\right]^T.
\end{split}
\end{equation}
$\Phi(\lambda)$ is the eigenfunction of the spin operator of a spin-$\frac{3}{2}$ baryons.
\begin{equation}
\begin{split}
\Phi\left(\frac{3}{2}\right)&=\left[1,0,0,0\right]^T,~
\Phi\left(\frac{1}{2}\right)=\left[0,1,0,0\right]^T,
\\
\Phi\left(-\frac{1}{2}\right)&=\left[0,0,1,0\right]^T,~
\Phi\left(-\frac{3}{2}\right)=\left[0,0,0,1\right]^T.
\end{split}
\end{equation}
With the above specific form, we can obtain the transition operator $S^{\mu}_{t}$.
\begin{equation}
\begin{split}
S^0_t=&0,~
S^x_t=\frac{1}{\sqrt{2}}\left[
\begin{array}{cccc}
-1 & 0 & \frac{1}{\sqrt{3}} & 0\\
0 & -\frac{1}{\sqrt{3}} & 0 & 1
\end{array}\right],
\\
S^y_t=&-\frac{i}{\sqrt{2}}\left[
\begin{array}{cccc}
1 & 0 & \frac{1}{\sqrt{3}} & 0\\
0 & \frac{1}{\sqrt{3}} & 0 & 1
\end{array}\right],~
S^x_t=\left[
\begin{array}{cccc}
0 & \sqrt{\frac{2}{3}} & 0 & 0\\
0 & 0 & \sqrt{\frac{2}{3}} & 0
\end{array}\right].
\end{split}
\end{equation}
The spin operator for the spin-$3\over 2$ particles can be derived
as $\bm{S}=\frac{3}{2}\bm{\sigma}_{rs}$, while
$\bm{\sigma}_{rs}\equiv-S^\dagger_{t\mu}\bm{\sigma}S^\mu_t$.

\section{Numerical results of the baryon-antibaryon system}\label{app_BBbar}

We calculate the possible molecular states formed by one baryon and
one antibaryon. Although the baryon-antibaryon systems may be not
stable, we give the possible molecular solutions for reference.

\subsection{The $B^{*}\bar{B}^{*}$ system}

We calculate the systems composed of one spin-$\frac{3}{2}$ baryon
and one spin-$\frac{3}{2}$ antibaryon. For the
$\Xi_{cc}^{*}\bar{\Omega}_{cc}^{*}$ and
$\bar{\Xi}_{cc}^{*}\Omega_{cc}^{*}$ systems, they are antiparticles
of each other and we only calculate the former system. We show the
binding energies and the root-mean-square radii of possible
molecular states in Tables~\ref{Table_B3Bbar3_01}
and~\ref{Table_B3Bbar3_23}. The mixing channels are the same as
those for the two spin-$\frac{3}{2}$ baryons systems. Therefore, the
angular momentum dependent operators are also the same.

\begin{table*}[htb]
	\centering
	\caption{The numerical results for the $B^*\bar{B}^*$ systems with $J=0,1$. $\Lambda $ is the cutoff parameter. ``$B.E.$" is the binding energy. $R_{rms}$ is the root-mean-square radius. $P_{i}$ is the percentage of the different channels.}\label{Table_B3Bbar3_01}
	\begin{tabular}{cccccccc}
		\hline
		States & $\Lambda$(MeV) & $B.E.$(MeV) & $R_{rms}$(fm) & $P_{S}(\%)$ & $P_{D1}(\%)$ & $P_{D2}(\%)$ & $P_{G}(\%)$\tabularnewline
		\hline
		$\left[\Xi_{cc}^{*}\bar{\Xi}_{cc}^{*}\right]_{J=0}^{I=0}$ & 940 & 1.96 & 2.96 & 81.9 & 18.1 &  & \tabularnewline
		& 960 & 4.45 & 2.28 & 75.7 & 24.3 &  & \tabularnewline
		& 980 & 8.78 & 1.85 & 70.5 & 29.5 &  & \tabularnewline
		\hline
		$\left[\Xi_{cc}^{*}\bar{\Xi}_{cc}^{*}\right]_{J=0}^{I=1}$ & 1000 & 2.68 & 1.82 & 99.3 & 0.7 &  & \tabularnewline
		& 1100 & 7.63 & 1.22 & 99.2 & 0.8 &  & \tabularnewline
		& 1200 & 14.88 & 0.95 & 99.1 & 0.9 &  & \tabularnewline
		\hline
		$\left[\Xi_{cc}^{*}\bar{\Omega}_{cc}^{*}\right]_{J=0}^{I=\frac{1}{2}}$ & 1400 & 6.40 & 1.19 & 99.7 & 0.3 &  & \tabularnewline
		& 1500 & 12.54 & 0.92 & 99.6 & 0.4 &  & \tabularnewline
		& 1600 & 20.92 & 0.76 & 99.5 & 0.5 &  & \tabularnewline
		\hline
		$\left[\Omega_{cc}^{*}\bar{\Omega}_{cc}^{*}\right]_{J=0}^{I=0}$ & 1250 & 1.48 & 2.32 & 98.5 & 1.5 &  & \tabularnewline
		& 1300 & 8.62 & 1.21 & 96.1 & 3.9 &  & \tabularnewline
		& 1350 & 27.87 & 0.81 & 93.9 & 6.1 &  & \tabularnewline
		\hline
		$\left[\Xi_{cc}^{*}\bar{\Xi}_{cc}^{*}\right]_{J=1}^{I=0}$ & 900 & 1.63 & 3.26 & 85.9 & 8.4 & 5.7 & 0.0\tabularnewline
		& 950 & 9.52 & 2.01 & 74.6 & 14.3 & 11.0 & 0.1\tabularnewline
		& 1000 & 27.92 & 1.52 & 67.7 & 16.6 & 15.6 & 0.1\tabularnewline
		\hline
		$\left[\Xi_{cc}^{*}\bar{\Xi}_{cc}^{*}\right]_{J=1}^{I=1}$ & 1000 & 1.07 & 2.71 & 99.0 & 0.8 & 0.2 & 0.0\tabularnewline
		& 1200 & 9.98 & 1.17 & 98.2 & 1.6 & 0.2 & 0.0\tabularnewline
		& 1400 & 27.05 & 0.82 & 97.6 & 2.1 & 0.3 & 0.0\tabularnewline
		\hline
		$\left[\Xi_{cc}^{*}\bar{\Omega}_{cc}^{*}\right]_{J=1}^{I=\frac{1}{2}}$ & 1400 & 4.18 & 1.44 & 99.7 & 0.2 & 0.1 & 0.0\tabularnewline
		& 1500 & 8.69 & 1.09 & 99.5 & 0.4 & 0.1 & 0.0\tabularnewline
		& 1600 & 14.97 & 0.89 & 99.3 & 0.6 & 0.1 & 0.0\tabularnewline
		\hline
		$\left[\Omega_{cc}^{*}\bar{\Omega}_{cc}^{*}\right]_{J=1}^{I=0}$ & 1250 & 1.67 & 2.26 & 98.5 & 1.0 & 0.5 & 0.0\tabularnewline
		& 1300 & 7.01 & 1.37 & 96.5 & 2.3 & 1.2 & 0.0\tabularnewline
		& 1350 & 18.68 & 1.02 & 94.4 & 3.4 & 2.2 & 0.0\tabularnewline
		\hline
	\end{tabular}
\end{table*}

\begin{table*}[htb]
	\centering
	\caption{The numerical results for the $B^*\bar{B}^*$ systems with $J=2,3$. $\Lambda $ is the cutoff parameter. ``$B.E.$" is the binding energy. $R_{rms}$ is the root-mean-square radius. $P_{i}$ is the percentage of the different channels.}\label{Table_B3Bbar3_23}
	\begin{tabular}{ccccccccc}
		\hline
		States & $\Lambda$(MeV) & $B.E.$(MeV) & $R_{rms}$(fm) & $P_{S}(\%)$ & $P_{D1}(\%)$ & $P_{D2}(\%)$ & $P_{G1}(\%)$ & $P_{G2}(\%)$\tabularnewline
		\hline
		$\left[\Xi_{cc}^{*}\bar{\Xi}_{cc}^{*}\right]_{J=2}^{I=0}$ & 900 & 1.15 & 3.39 & 92.4 & 1.9 & 5.7 & 0.0 & \tabularnewline
		& 950 & 6.50 & 2.00 & 87.4 & 3.0 & 9.5 & 0.1 & \tabularnewline
		& 1000 & 18.13 & 1.51 & 83.9 & 3.8 & 12.1 & 0.2 & \tabularnewline
		\hline
		$\left[\Xi_{cc}^{*}\bar{\Xi}_{cc}^{*}\right]_{J=2}^{I=1}$ & 1400 & 3.80 & 1.69 & 98.3 & 0.2 & 1.5 & 0.0 & \tabularnewline
		& 1600 & 10.21 & 1.18 & 97.8 & 0.2 & 2.0 & 0.0 & \tabularnewline
		& 1800 & 19.13 & 0.94 & 97.2 & 0.3 & 2.5 & 0.0 & \tabularnewline
		\hline
		$\left[\Xi_{cc}^{*}\bar{\Omega}_{cc}^{*}\right]_{J=2}^{I=\frac{1}{2}}$ & 1600 & 4.31 & 1.46 & 99.6 & 0.1 & 0.3 & 0.0 & \tabularnewline
		& 1800 & 9.93 & 1.07 & 99.3 & 0.1 & 0.6 & 0.0 & \tabularnewline
		& 2000 & 17.62 & 0.87 & 98.9 & 0.1 & 1.0 & 0.0 & \tabularnewline
		\hline
		$\left[\Omega_{cc}^{*}\bar{\Omega}_{cc}^{*}\right]_{J=2}^{I=0}$ & 1300 & 4.64 & 1.55 & 97.9 & 0.4 & 1.7 & 0.0 & \tabularnewline
		& 1400 & 18.22 & 1.04 & 94.8 & 1.2 & 4.0 & 0.0 & \tabularnewline
		& 1500 & 45.21 & 0.84 & 90.9 & 2.3 & 6.7 & 0.1 & \tabularnewline
		\hline
		$\left[\Xi_{cc}^{*}\bar{\Xi}_{cc}^{*}\right]_{J=3}^{I=0}$ & 800 & 7.26 & 1.59 & 96.7 & 0.7 & 2.5 & 0.0 & 0.1\tabularnewline
		& 900 & 17.90 & 1.29 & 94.7 & 1.2 & 3.9 & 0.0 & 0.2\tabularnewline
		& 1000 & 36.68 & 1.19 & 89.2 & 2.4 & 8.0 & 0.1 & 0.3\tabularnewline
		\hline
		$\left[\Xi_{cc}^{*}\bar{\Xi}_{cc}^{*}\right]_{J=3}^{I=1}$ & 2200 & 2.16 & 2.27 & 96.9 & 0.2 & 2.9 & 0.0 & 0.0\tabularnewline
		& 2400 & 3.76 & 1.86 & 96.1 & 0.3 & 3.6 & 0.0 & 0.0\tabularnewline
		& 2600 & 5.78 & 1.59 & 95.3 & 0.3 & 4.4 & 0.0 & 0.0\tabularnewline
		\hline
		$\left[\Xi_{cc}^{*}\bar{\Omega}_{cc}^{*}\right]_{J=3}^{I=\frac{1}{2}}$ & 1800 & 1.69 & 2.21 & 99.4 & 0.1 & 0.5 & 0.0 & 0.0\tabularnewline
		& 2000 & 3.32 & 1.71 & 99.0 & 0.1 & 0.9 & 0.0 & 0.0\tabularnewline
		& 2200 & 5.33 & 1.45 & 98.5 & 0.1 & 1.4 & 0.0 & 0.0\tabularnewline
		\hline
		$\left[\Omega_{cc}^{*}\bar{\Omega}_{cc}^{*}\right]_{J=3}^{I=0}$ & 1200 & 1.15 & 2.56 & 99.3 & 0.1 & 0.6 & 0.0 & 0.0\tabularnewline
		& 1300 & 5.06 & 1.53 & 97.6 & 0.4 & 2.0 & 0.0 & 0.0\tabularnewline
		& 1400 & 14.06 & 1.17 & 94.5 & 0.9 & 4.6 & 0.0 & 0.0\tabularnewline
		\hline
	\end{tabular}
\end{table*}

\begin{figure*}
	\centering
	\subfigure[$V_{11}$]{
		\label{fig:vxxbar0011}
		\includegraphics[width=0.31\linewidth]{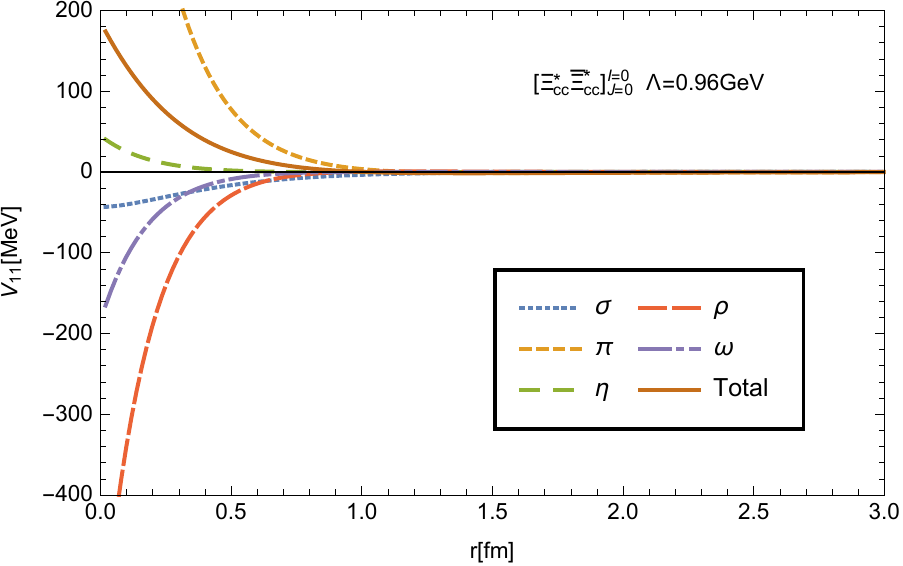}
	}
	\subfigure[$V_{12}$]{
		\label{fig:vxxbar0012}
		\includegraphics[width=0.31\linewidth]{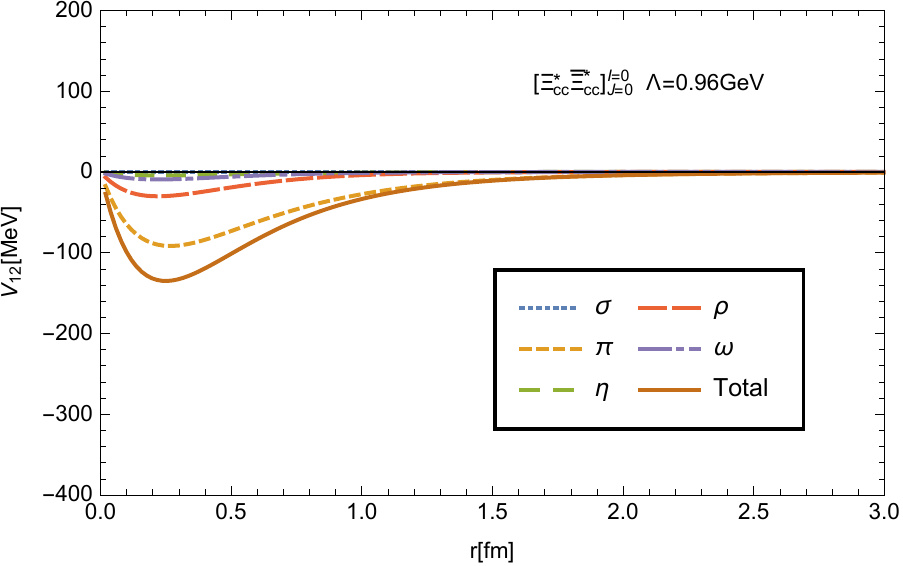}
	}
	\subfigure[$V_{22}$]{
		\label{fig:vxxbar0022}
		\includegraphics[width=0.31\linewidth]{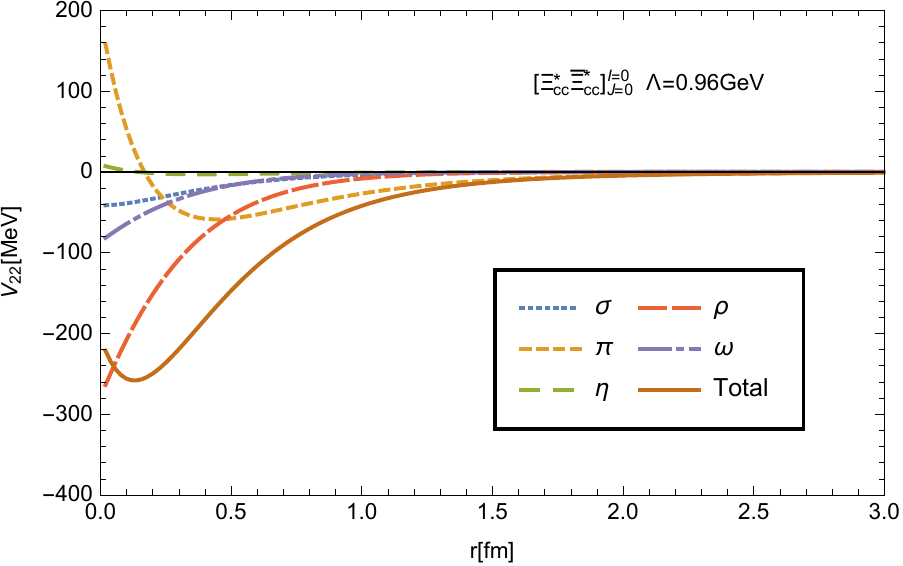}
	}
	\caption{The interactions potentials for the system $\left[\Xi_{cc}^*\Xi_{cc}^*\right]^{I=1}_{J=2}$. $V_{11}$, $V_{12}$ and $V_{22}$ denote the ${}^{1}S_{0}\leftrightarrow{}^{1}S_{0}$, ${}^{1}S_{0} \leftrightarrow{}^{5}D_{0}$ and ${}^{5}D_{0}\leftrightarrow{}^{5}D_{0}$ transitions potentials.}
	\label{potential:XXbar00}
\end{figure*}

For the $\left[\Xi_{cc}^{*}\bar{\Xi}_{cc}^{*}\right]_{J=0}^{I=0}$
system, we find a binding solution with binding energy 1.96-8.78 MeV
when we choose the cutoff parameter between 0.94 GeV and 0.98 GeV.
The $D$-wave contribution is significant and increases with the
cutoff parameter. When we choose the cutoff around 1.0 GeV, the
percentage of the ${}^5D_0$ channel is over 30\%. We give the
interaction potentials of the system in
Fig.~\ref{potential:XXbar00}. If only the $S$-wave is considered,
the barely attractive potential could not produce a bound state. But
the couple-channel effect from an adequately attractive $D$-wave
makes it possible. For the
$\left[\Xi_{cc}^{*}\bar{\Xi}_{cc}^{*}\right]_{J=0}^{I=1}$ system,
the binding energy is 2.68-14.88 MeV while the cutoff is 1.0-1.2
GeV. For the
$\left[\Xi_{cc}^{*}\bar{\Omega}_{cc}^{*}\right]_{J=0}^{I=\frac{1}{2}}$
system, we obtain a bound state. The binding energy of the system is
12.54 MeV when the cutoff parameter is 1.5 GeV. Although the
root-mean-square radius of the system is 0.76fm when we choose a
large cutoff parameter, the system still seems to be a molecular
states with the cutoff parameter less than 1.5 GeV. For the
$\left[\Omega_{cc}^{*}\bar{\Omega}_{cc}^{*}\right]_{J=0}^{I=0}$
system, a binding solution with the binding energy 1.48-27.87 MeV
appears with the cutoff parameter from 1.25 GeV to 1.35 GeV.

For the $\left[\Xi_{cc}^{*}\bar{\Xi}_{cc}^{*}\right]_{J=1}^{I=0}$
system, we obtain a bound state with the binding energy 1.63-27.92
MeV while the cutoff parameter is 0.9-1.0 GeV. The $G$-wave
contribution of the system is almost zero. However, the total
contribution of the ${}^3D_1$ and ${}^7D_1$ channel is about 30\%.
For the other three systems with total spin 1,
$\left[\Xi_{cc}^{*}\bar{\Xi}_{cc}^{*}\right]_{J=1}^{I=1}$,
$\left[\Xi_{cc}^{*}\bar{\Omega}_{cc}^{*}\right]_{J=1}^{I=\frac{1}{2}}$
and $\left[\Omega_{cc}^{*}\bar{\Omega}_{cc}^{*}\right]_{J=1}^{I=0}$,
we all find binding solutions with reasonable binding energies and
root-mean-square radii. The dominant parts of their wave functions
are all $S$-wave.

For the systems
$\left[\Xi_{cc}^{*}\bar{\Xi}_{cc}^{*}\right]_{J=2}^{I=0}$, a bound
state appears with binding energy about 1.15 MeV, when the cutoff
parameter is around 0.95 GeV. The $D$-waves are important for the
system. The contribution of the ${}^5D_2$ channel is almost 10\%
when the cutoff parameter is 0.95 GeV. For the
$\left[\Xi_{cc}^{*}\bar{\Xi}_{cc}^{*}\right]_{J=2}^{I=1}$,
$\left[\Xi_{cc}^{*}\bar{\Omega}_{cc}^{*}\right]_{J=2}^{I=\frac{1}{2}}$
and $\left[\Omega_{cc}^{*}\bar{\Omega}_{cc}^{*}\right]_{J=2}^{I=0}$
systems, each of them has a loosely bound state solution, and
dominant $S$-wave part in the total wave function. We show the
binding energies, root-mean-square radii as well as the
contributions of all the channels in the
Table~\ref{Table_B3Bbar3_23}.

For the $\left[\Xi_{cc}^{*}\bar{\Xi}_{cc}^{*}\right]_{J=3}^{I=0}$
system, we obtain a bound state with binding energy 7.26-36.68 MeV
when the cutoff varies from 0.8 GeV to 1.0 GeV. For the
$\left[\Xi_{cc}^{*}\bar{\Xi}_{cc}^{*}\right]_{J=3}^{I=1}$ system,
the loosely bound state solution appears until we change the cutoff
parameter over 2.0 GeV. We change the cutoff parameter from 2.2 GeV
to 2.6 GeV, and the binding energy increases from 2.16 MeV to 5.78
MeV. For the
$\left[\Xi_{cc}^{*}\bar{\Omega}_{cc}^{*}\right]_{J=3}^{I=\frac{1}{2}}$
system, we find a binding solution with binding energy 1.69-5.53 MeV
when the cutoff parameter is 1.8-2.2 GeV. The binding energies of
the above two systems are stable for the cutoff parameter in the
range over 2.0 GeV. For the
$\left[\Omega_{cc}^{*}\bar{\Omega}_{cc}^{*}\right]_{J=3}^{I=0}$
system, a bound state with binding energy 1.15-14.06 MeV appears
when the cutoff parameter changes from 1.2 GeV to 1.4 GeV.

For the possible systems composed of one spin-$\frac{3}{2}$ baryon
and one spin-$\frac{3}{2}$ antibaryon, they are all good candidates
of molecular states. A baryon-antibaryon system may be unstable
especially when the threshold of three mesons is open. Some of these
molecular systems may appear as an enhancement in the baryon and
antibaryon invariant mass spectrum, or a narrow resonance state etc.

\subsection{The $B^{*}\bar{B}$ system}

We also consider the systems composed of one spin-$\frac{3}{2}$
baryon and one spin-$\frac{1}{2}$ antibaryon, $B^{*}\bar{B}$. The
mixing channels are the same as that for the two baryon systems with
the same total angular momentum. The numerical results are shown in
Table~\ref{Table_B3Bbar1}.

\begin{table*}[htb]
	\centering
	\caption{The numerical results for the $B^*\bar{B}$ systems. $\Lambda $ is the cutoff parameter. ``$B.E.$" is the binding energy. $R_{rms}$ is the root-mean-square radius. $P_{i}$ is the percentage of the different channels.}\label{Table_B3Bbar1}
	\begin{tabular}{cccccccc}
		\hline
		States & $\Lambda$(MeV) &  $B.E.$(MeV) & $R_{rms}$(fm) & $P_{S}(\%)$ & $P_{D1}(\%)$ & $P_{D2}(\%)$ & $P_{G}(\%)$\tabularnewline
		\hline
		$\left[\Xi_{cc}^{*}\bar{\Xi}_{cc}\right]_{J=1}^{I=0}$ & 900 & 1.67 & 2.21 & 99.8 & 0.1 & 0.1 & \tabularnewline
		& 1000 & 8.82 & 1.24 & 99.6 & 0.2 & 0.2 & \tabularnewline
		& 1100 & 18.21 & 1.03 & 99.3 & 0.4 & 0.3 & \tabularnewline
		\hline
		$\left[\Xi_{cc}^{*}\bar{\Xi}_{cc}\right]_{J=1}^{I=1}$ & 2500 & 2.96 & 1.79 & 99.3 & 0.1 & 0.6 & \tabularnewline
		& 3000 & 5.68 & 1.39 & 99.0 & 0.1 & 0.9 & \tabularnewline
		& 3500 & 8.67 & 1.19 & 98.7 & 0.1 & 1.2 & \tabularnewline
		\hline
		$\left[\Xi_{cc}^{*}\bar{\Omega}_{cc}\right]_{J=1}^{I=\frac{1}{2}}$ & 1800 & 1.82 & 2.06 & 99.9 & 0.0 & 0.1 & \tabularnewline
		& 2200 & 4.86 & 1.39 & 99.8 & 0.0 & 0.2 & \tabularnewline
		& 2600 & 8.06 & 1.15 & 99.7 & 0.0 & 0.3 & \tabularnewline
		\hline
		$\left[\Omega_{cc}^{*}\bar{\Xi}_{cc}\right]_{J=1}^{I=\frac{1}{2}}$ & 1800 & 1.86 & 2.05 & 99.9 & 0.0 & 0.1 & \tabularnewline
		& 2200 & 4.93 & 1.39 & 99.8 & 0.0 & 0.2 & \tabularnewline
		& 2600 & 8.15 & 1.15 & 99.7 & 0.0 & 0.3 & \tabularnewline
		\hline
		$\left[\Omega_{cc}^{*}\bar{\Omega}_{cc}\right]_{J=1}^{I=0}$ & 1300 & 1.78 & 2.09 & 99.8 & 0.0 & 0.2 & \tabularnewline
		& 1500 & 8.32 & 1.22 & 99.1 & 0.0 & 0.9 & \tabularnewline
		& 1700 & 19.23 & 0.97 & 97.6 & 0.1 & 2.3 & \tabularnewline
		\hline
		$\left[\Xi_{cc}^{*}\bar{\Xi}_{cc}\right]_{J=2}^{I=0}$ & 1000 & 2.52 & 2.26 & 96.3 & 0.9 & 2.8 & 0.0\tabularnewline
		& 1040 & 10.77 & 1.39 & 94.6 & 1.3 & 4.1 & 0.0\tabularnewline
		& 1080 & 26.29 & 1.04 & 93.7 & 1.4 & 4.9 & 0.0\tabularnewline
		\hline
		$\left[\Xi_{cc}^{*}\bar{\Xi}_{cc}\right]_{J=2}^{I=1}$ & 1400 & 1.78 & 2.17 & 99.6 & 0.1 & 0.3 & 0.0\tabularnewline
		& 1600 & 5.88 & 1.36 & 99.4 & 0.2 & 0.4 & 0.0\tabularnewline
		& 1800 & 11.66 & 1.05 & 99.2 & 0.3 & 0.5 & 0.0\tabularnewline
		\hline
		$\left[\Xi_{cc}^{*}\bar{\Omega}_{cc}\right]_{J=2}^{I=\frac{1}{2}}$ & 1500 & 1.70 & 2.10 & 99.9 & 0.0 & 0.1 & 0.0\tabularnewline
		& 1700 & 5.26 & 1.32 & 99.8 & 0.1 & 0.1 & 0.0\tabularnewline
		& 1900 & 10.26 & 1.02 & 99.7 & 0.1 & 0.2 & 0.0\tabularnewline
		\hline
		$\left[\Omega_{cc}^{*}\bar{\Xi}_{cc}\right]_{J=2}^{I=\frac{1}{2}}$ & 1500 & 1.74 & 2.08 & 99.9 & 0.0 & 0.1 & 0.0\tabularnewline
		& 1700 & 5.34 & 1.32 & 99.8 & 0.1 & 0.1 & 0.0\tabularnewline
		& 1900 & 10.38 & 1.02 & 99.7 & 0.1 & 0.2 & 0.0\tabularnewline
		\hline
		$\left[\Omega_{cc}^{*}\bar{\Omega}_{cc}\right]_{J=2}^{I=0}$ & 1300 & 2.55 & 1.83 & 99.6 & 0.1 & 0.3 & 0.0\tabularnewline
		& 1400 & 11.71 & 1.05 & 98.8 & 0.3 & 0.9 & 0.0\tabularnewline
		& 1500 & 30.91 & 0.77 & 97.8 & 0.5 & 1.7 & 0.0\tabularnewline
		\hline
	\end{tabular}
\end{table*}

For the $\left[\Xi_{cc}^{*}\bar{\Xi}_{cc}\right]_{J=1}^{I=0}$
system, we find a bound state with binding energy 1.67-18.21 MeV
when the cutoff parameter is 0.9-1.1 GeV. For the
$\left[\Xi_{cc}^{*}\bar{\Xi}_{cc}\right]_{J=1}^{I=1}$ system, a very
loosely bound state solution with binding energy 2.96 MeV appears
with the cutoff parameter is 2.5 GeV. The bound state is quite
insensitive to the cutoff parameter. The $D$-wave contribution is
very small. For the
$\left[\Xi_{cc}^{*}\bar{\Omega}_{cc}\right]_{J=1}^{I=\frac{1}{2}}$
system, we find a binding solution insensitive to the cutoff
parameter. The binding energy increases from 1.82 MeV to 8.06 MeV
along with the cutoff parameter changing from 1.8 GeV to 2.6 GeV.
The
$\left[\Omega_{cc}^{*}\bar{\Xi}_{cc}\right]_{J=1}^{I=\frac{1}{2}}$
system is quite similar to the
$\left[\Xi_{cc}^{*}\bar{\Omega}_{cc}\right]_{J=1}^{I=\frac{1}{2}}$
system. A bound state with binding energy 1.86-8.15 MeV appears with
the same range of the cutoff parameter. It is not surprising because
the
$\left[\Omega_{cc}^{*}\bar{\Xi}_{cc}\right]_{J=1}^{I=\frac{1}{2}}$
state and the
$\left[\bar{\Xi}_{cc}^{*}\Omega_{cc}\right]_{J=1}^{I=\frac{1}{2}}$
state are degenerate in the heavy quark limit. For the
$\left[\Omega_{cc}^{*}\bar{\Omega}_{cc}\right]_{J=1}^{I=0}$ system,
we obtain a binding solution with binding energy 1.78-19.23 MeV when
the cutoff parameter is 1.3-1.7 GeV. The $S$-wave contribution is
99\%.

The $G$-wave contributions for the $J=2$ systems are negligible. The
contributions of $D$-waves are also very small. For the
$\left[\Xi_{cc}^{*}\bar{\Xi}_{cc}\right]_{J=2}^{I=0}$ system, we
obtain a binding solution with binding energy 2.52-26.29 MeV when we
choose the cutoff parameter from 1.0 GeV to 1.08 GeV. For the
$\left[\Xi_{cc}^{*}\bar{\Xi}_{cc}\right]_{J=2}^{I=1}$ system, we
find a loosely bound state. When the cutoff parameter changes from
1.4 GeV to 1.8 GeV, the binding energy increases from 1.78 MeV to
11.66 MeV. For the
$\left[\Xi_{cc}^{*}\bar{\Omega}_{cc}\right]_{J=2}^{I=\frac{1}{2}}$
and
$\left[\Omega_{cc}^{*}\bar{\Xi}_{cc}\right]_{J=2}^{I=\frac{1}{2}}$
systems, we obtain binding solutions when the cutoff parameter
changes from 1.5 GeV to 1.9 GeV. The binding energy of the
$\Xi_{cc}^{*}\bar{\Omega}_{cc}$ system is 1.70-10.26 MeV, while that
of the $\Omega_{cc}^{*}\bar{\Xi}_{cc}$ system is 1.74-10.38 MeV. For
the $\left[\Omega_{cc}^{*}\bar{\Omega}_{cc}\right]_{J=2}^{I=0}$
system, a bound state with binding energy 2.55-30.91 MeV appears
when the cutoff parameter is 1.3-1.5 GeV.

For the four $J=1$ systems,
$\left[\Xi_{cc}^{*}\bar{\Xi}_{cc}\right]_{J=1}^{I=1}$,
$\left[\Xi_{cc}^{*}\bar{\Omega}_{cc}\right]_{J=1}^{I=\frac{1}{2}}$,
$\left[\Omega_{cc}^{*}\bar{\Xi}_{cc}\right]_{J=1}^{I=\frac{1}{2}}$
and $\left[\Omega_{cc}^{*}\bar{\Omega}_{cc}\right]_{J=1}^{I=0}$,
their loosely bound solutions are very insensitive to the cutoff
parameter.

\end{appendix}

\newpage

\bibliographystyle{spmpsci}

\bibliography{ref}

\end{document}